\documentclass[a4paper,onecolumn,11pt,unpublished]{quantumarticle}
\pdfoutput=1
\usepackage[utf8]{inputenc} 
\usepackage{hyperref}       
\usepackage{url}            
\usepackage{booktabs}       
\usepackage{amsfonts}       
\usepackage{xcolor}         
\usepackage{xurl}           
\usepackage{amssymb}
\usepackage{amsmath}
\usepackage{mathtools}
\usepackage{placeins} 

\newcommand{\red}[1]{\textcolor{red}{#1}}

\usepackage[
sorting=none,
backref=true,
url=false,
isbn=false,
backend=biber,
style=numeric-comp,
]{biblatex}
\usepackage{hyperref}
\hypersetup{
     colorlinks   = true,
     citecolor    = blue,
     linkcolor    = blue,
}

\DeclareMathOperator*{\argmax}{arg\,max}
\DeclareMathOperator*{\argmin}{arg\,min}
\newcommand{\I}{\mathbb{I}} 
\newcommand{\E}{\mathbb{E}} 
\DeclarePairedDelimiter\ceil{\lceil}{\rceil}

\newcommand{\mD}{\mathcal{D}}

\newcommand{\mL}{\mathcal{L}}

\newcommand{\mP}{\mathcal{P}}

\newcommand{\mT}{\mathcal{T}}

\newcommand{\mX}{\mathcal{X}}
\newcommand{\mY}{\mathcal{Y}}

\newcommand{\bR}{\mathbb{R}} 

\DeclareMathAlphabet{\mathpzc}{OT1}{pzc}{m}{it}

\newcommand{\norm}[1]{\lVert#1\rVert}

\newcommand{\mPhat}{\hat{\mathcal{P}}} 
\newcommand{\Nhat}{\hat{N}} 
\newcommand{\mS}{\mathcal{S}} 
\newcommand{\mE}{\mathcal{\mathcal{E}}} 
\DeclareMathOperator{\wt}{wt} 

\DeclareMathAlphabet\mathbfcal{OMS}{cmsy}{b}{n}

\newcommand{\R}{\mathbb{R}} 

\usepackage{amsthm}
\theoremstyle{definition}

\newtheorem{theorem}{Theorem}
\newtheorem*{theorem*}{Theorem}

\newtheorem*{lemma*}{Lemma}
\newtheorem{corollary}[theorem]{Corollary}
\newtheorem*{corollary*}{Corollary}
\newtheorem{proposition}[theorem]{Proposition}

\DeclareMathOperator{\DDD}{DDD}
\DeclareMathOperator{\DDDD}{Detector\DDD}
\DeclareMathOperator{\MLD}{MLD}
\DeclareMathOperator{\poly}{poly}

\begin{document}

\title{Importance sampling for data-driven decoding of quantum error-correcting codes}

\author{Evan Peters}
\thanks{\email{e6peters@uwaterloo.ca}{e6peters@uwaterloo.ca}}
\affiliation{Department of Applied Mathematics, University of Waterloo, Waterloo, ON N2L 3G1}
\orcid{0000-0001-7083-6733}

\maketitle

\begin{abstract}
Data-driven decoding ($\DDD$) -- learning to decode syndromes of (quantum) error-correcting codes by learning from data -- can be a difficult problem due to several atypical and poorly understood properties of the training data. We introduce a theory of \textit{example importance} that clarifies these unusual aspects of $\DDD$: For instance, we show that $\DDD$ of a simple error-correcting code is equivalent to a noisy, imbalanced binary classification problem. We show that an existing importance sampling technique of training neural decoders on data generated with higher error rates introduces a tradeoff between class imbalance and label noise. We apply this technique to show robust improvements in the accuracy of neural network decoders trained on syndromes sampled at higher error rates, and provide heuristic arguments for finding an optimal error rate for training data. We extend these analyses to decoding quantum codes involving multiple rounds of syndrome measurements, suggesting broad applicability of both example importance and turning the knob for improving experimentally relevant data-driven decoders. 
\end{abstract}

\section{Introduction}

Data-driven decoding for error-correcting codes poses a peculiar learning problem: You might run a large error correction experiment to generate millions of decoding examples, only to find that most of these training data are useless. Regardless, you push ahead and manage to train a neural network that achieves 99.99\% test accuracy, but later realize that this is \textit{strikingly bad} performance for the task at hand. The challenge for neural decoders -- algorithms that learn how to map syndromes of a (quantum) error-correcting code into correction operations to restore the encoded information -- is that most syndromes correspond to low-weight errors, which existing baseline decoders can easily handle. To be useful, a neural decoder must outperform baseline decoders for syndromes that might appear once (or never) in a large training dataset. Despite this, neural decoders are demonstrably better than state-of-the-art baseline decoders in some classical and quantum error correction experiments \cite{bausch_learning_2023,KimJRKOV18,pmlr-v130-garcia-satorras21a,choukroun_error_2022,kwak2023boosting,senior2025scalablerealtimeneuraldecoder}. Data-driven decoding therefore deserves special attention to better understand its peculiarities and design techniques to further improve neural decoders.  

\begin{figure}[ht]
    \centering
    \includegraphics[width=\linewidth]{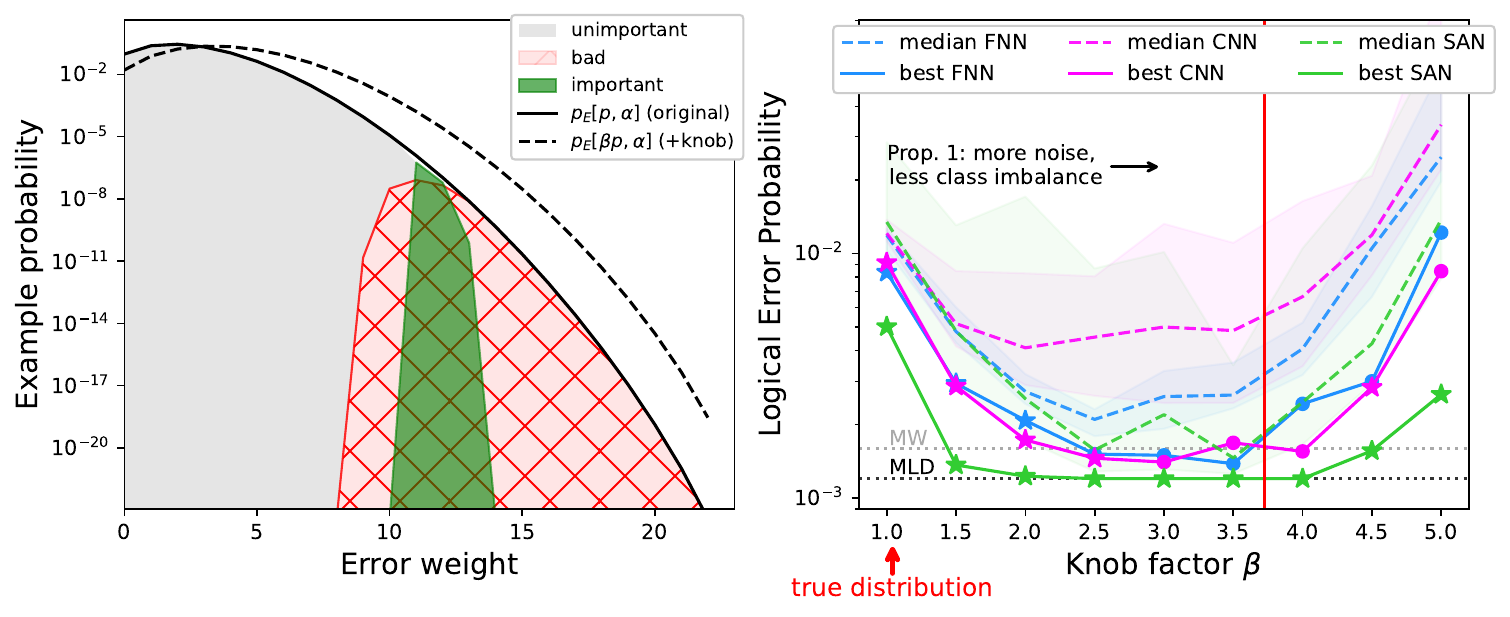}
    \caption{\textbf{(left) Important examples in the repetition code have weight close to $n/2$, and are much rarer than unimportant examples.  \textit{Turning the knob} increases a training dataset's support on both important examples and bad examples}. An example error distribution $p_E[p=0.15, \alpha=0.33]$ with $n=22$ is mostly supported on \textit{unimportant} errors that are correctly decoded by a minimum-weight baseline decoder. The dashed line shows that the training distribution $p_E[\beta p, \alpha]$ has greater support on important examples after \textit{turning the knob}. \textbf{(right) Turning the knob robustly improves neural decoder performance.}  
    Each point (dash) represents the best (median) test error from 150 repetition code $\DDD$ experiments per $\beta$ with $N=2000$ and different initializations and hyperparameters, for three different architectures ($\sim 4000$ neural decoders total). Shading denotes IQ range and $\star$ denotes outperforming a lookup table on the training dataset. The vertical red line predicts when turning the knob becomes harmful (Eq.~\ref{eq:threshold}), resulting in a ``U''-shaped graph of logical error probability vs $\beta$.}
    \label{fig:rep_code_importance}
\end{figure}

In this work, we first formalize \textit{data-driven decoding} ($\DDD$), the problem of learning to decode from training examples. We develop a theory of \textit{example importance} that explains challenges that are seemingly unique to $\DDD$ in terms of more standard machine learning phenomena. This theory motivates a rarely used data augmentation technique for improving neural decoders: \textit{turning the knob} to increase the dataset error rate. We argue that the mechanism behind this technique is to generate additional important examples and that in a simple decoding problem this technique results in a tradeoff between label noise and class imbalance. Fig.~\ref{fig:rep_code_importance} depicts example importance in training data and how/when turning the knob improves accuracy for $\DDD$ of a classical repetition code.

After characterizing $\DDD$ for a simple decoding problem, we show that these empirical observations hold for more complex decoding problems involving quantum error-correcting codes (QECCs). We provide empirical evidence that this technique robustly improves decoders for multiple rounds of syndrome measurement. We show that multiple rounds of decoding with a single logical qubit is formally equivalent to an imbalanced, binary classification problem, and we propose a heuristic for finding a good error rate for generating training data. The broad applicability of both sample importance and turning the knob suggests room for improvements in both our theoretical understanding and experimental implementations of neural decoders.

\section{Decoding quantum error-correcting codes}
While promising, quantum computers are prone to failure due to device imperfections, thermal noise, and spurious coupling with their environment. It is widely accepted that useful quantum computing will require error correction, which involves encoding relatively fragile quantum information into a higher-dimensional system that is more robust to errors. We focus on stabilizer QECCs: Let $\mPhat_n := \langle\{I, X, Y, Z\}^{\otimes n} \rangle / \{\pm I, \pm iI\}$ be the $n$-qubit Pauli group mod phase, and define the \textit{weight} $\wt(e)$ of $e \in \mPhat_n$ to be the number of non-identity terms in $e$. An $[[n, k, d]]$ stabilizer code defines a partitioning of $\mPhat_n$ into $\mL \times \mS \times \mT$, where $\mS$ denotes the stabilizer group generated by $n-k$ commuting elements of $\mPhat_n$ with $-I \notin \mS$, $\mL$ is the centralizer (mod $\{\pm I, \pm iI\}$) of $\mS$  with $\min_{x\in \mL: x\neq I} \wt(x) :=d$, and $\mT$ is the span of $n-k$ destabilizer generators each anticommuting with exactly one generator of $\mS$. Thus,  $|\mL| = 2^{2k}$, $|\mS| = |\mT| = 2^{n-k}$, and every $e \in \mPhat_n$ is uniquely written $e = \ell(e) s(e) t(e)$ for $\ell(e) \in \mL$, $s(e) \in \mS$, and $t(e)\in\mT$. See Appendix~\ref{app:background} for additional background.

In the context of stabilizer codes, an error in a quantum device can be modeled by sampling an operator $e \in \mPhat_n$ from some distribution of errors $p_E: \mPhat_n \rightarrow \R^+$. The error's \textit{syndrome} $\sigma(e) \in \{0,1\}^{n-k}$ is a bitstring description for $t(e)$ that can be determined via measurements. A characteristic of QECCs is that the errors in $\mS$ do not affect the encoded quantum information, so $\ell(e)$ is the only salient component of $e$. Thus, the goal of \textbf{degenerate decoding} of a QECC is to determine $\ell(e)$ given $\sigma$. However, for some codes (e.g. classical codes) the more relevant task of \textbf{non-degenerate decoding} is to determine $e$ given $\sigma$. If we denote a set of \textit{labels} as $\mY$, then for degenerate decoding $\mY=\{0,1\}^{2k}$ specifies all possible values of $y(e):=\ell(e) \in \mL$, for nondegenerate decoding $\mY = \{0,1\}^{2n}$ specifies possible errors $y(e)=e \in \mPhat_n$, and for classical codes $\mY=\{0,1\}^n$ is sufficient to represent bitflip errors in $\langle \{I, X\}^{\otimes n}\rangle$. Either way, we write the conditional distribution of $y(e)$ given $\sigma(e)$ induced by $p_E$ as $p_{Y|\Sigma}$. A maximum likelihood decoder ($\MLD$) is any function that outputs the most likely value for $y$ given $\sigma$, and does either degenerate or non-degenerate decoding depending on the context. The $\MLD$ may not be unique; for each code and $p_E$ we denote a fixed choice of $\MLD$ by $f^*$ satisfying
\begin{align}
    f^*(\sigma) &\in \argmax_{y \in \mY} p_{Y|\Sigma} (y|\sigma).
\end{align}

\subsection{Data-driven decoding}

We can now formalize data-driven decoding ($\DDD$) as a learning problem. An instance of $\DDD$ is specified by a particular code along with a dataset $D_N$ of $N$ \textit{examples}, i.e. (syndrome, label) pairs $(\sigma, y)$. Each label is either the physical error $e$ or logical error $\ell(e)$ that gave rise to the syndrome. The distribution of $(\sigma, y)$ pairs is induced by an error distribution $p_E$ (typically over $\mPhat_n$) describing a physical error model for some device. $D_N$ is therefore generated by (i) sampling errors $(e_1, \dots, e_N) \sim p_E^{\otimes N}$ and then (ii) computing each example $(\sigma(e_i), y(e_i))$. The dataset is then given by $D_N := \{(\sigma_1, y_1), \dots, (\sigma_N, y_N)\}$. An \textit{example} is a pair $(\sigma_i, y_i)$ sampled according to the distribution $\mD: \{0,1\}^{n-k} \times \mathcal{Y} \rightarrow \R^+$, and a \textit{sample} (e.g. $D_N$) is any set of examples.

Given an instance of $\DDD$, the goal is to learn a decoding rule $\hat{f}: \{0,1\}^{n-k} \rightarrow \mathcal{Y}$ that decodes syndromes of errors sampled from $p_E$ with small \textit{logical error probability} (LEP) $\Pr_{(\sigma, y)\sim \mD}(\hat{f}(\sigma) \neq y)$. This is typically done by minimizing some loss function $L: \mY \times \mY \rightarrow \bR$, e.g. selecting a model $\hat{f}$ that minimizes $\sum_{i=1}^N L(\hat{f}(\sigma_i), y_i)$. To be useful, $\hat{f}$ should have lower LEP than a reasonable choice of baseline $f_0$. A \textit{neural decoder} is any $\hat{f}$ implemented using a neural network optimized with respect to the loss $L$, but for brevity we will use neural decoder to refer to any data-driven decoder. $\DDD$ is frequently studied in the literature of data-driven decoders: The nondegenerate variant was studied in Refs.~\cite{torlai_neural_2017,krastanov_deep_2017,varsamopoulos_decoding_2018,varsamopoulos_comparing_2020,breuckmann_scalable_2018,ni_neural_2020,davaasuren_general_2020}, and the degenerate variant was studied in Refs.~\cite{maskara_advantages_2019,davaasuren_general_2020,wagner_symmetries_2020, bhoumikEfficientDecodingSurface2021,overwater_neural_network_2022,meinerz_scalable_2022,cao_qecgpt_2023,gicev_scalable_2023,wang_transformer_qec_2023,egorovENDEquivariantNeural2023,Choukroun_Wolf_2024}, and is also applicable to \textit{generative} modeling, such as Refs.~\cite{krastanov_deep_2017,varsamopoulos_decoding_2018,varsamopoulos_comparing_2020,cao_qecgpt_2023}. $\DDD$ is a simplified version of learning to decode QECCs, since the $(\sigma, y)$ examples cannot actually be produced by experiments. Later in Sec.~\ref{sec:detectorddd} we will investigate an extension of $\DDD$ studied in Refs.~\cite{chamberland_deep_2018,baireuther_machine_2018,varsamopoulos2019decodingsurfacecodedistributed,
zhang_scalable_2023,varbanov_neural_2023,chamberlandTechniquesCombiningFast2023,lange_data_driven_2023,bausch_learning_2023,PhysRevResearch.6.L032004} that better describes data-driven decoding using datasets that can be sampled from experimental devices.

Solving $\DDD$ optimally given a polynomial amount of data is worst-case computationally intractable. This follows from the established hardness of optimally decoding QECCs \cite{hsiehNPhardnessDecodingQuantum2011,kuo_lu_2012,iyerHardnessDecodingQuantum2013} and the observation that a dataset $D_N$ with $N \sim \poly(n)$ can be generated in $\poly(n)$ time given knowledge of $p_E$ (e.g. via Gaussian elimination), implying that any $\poly(n)$-time decoding algorithm provided with $\poly(n)$ data can be simulated by some other $\poly(n)$-time decoder. Conversely, an exponential amount of data can, after an initial pre-processing stage, enable a constant-time algorithm (i.e. hash table) to perform near optimally. Thus, optimally solving $\DDD$ is too ambitious and here we are just interested in practical techniques for designing and implementing neural network decoders.

\section{Example importance for data-driven decoders}

In $\DDD$, some training examples are more important than others. a neural decoder is only relevant to the extent that it correctly decodes syndromes that an existing baseline decoder $f_0$ decodes incorrectly, but only some examples are useful for this purpose. Given a fixed choice of $f_0$, we classify examples as follows:
\begin{itemize}
    \item A \textit{good example} is any $(\sigma, y)$ for which $y = f^*(\sigma)$ (for a particular $\MLD$ $f^*$). By definition, any $\MLD$ is a lookup table on the set of all good examples. The total probability of good examples is $\Pr(\text{good}):=\Pr_{(\sigma, y)}(y=f^*(\sigma))$.
    \item A \textit{bad example} is any $(\sigma, y)$ that is not a good example. For instance, in the nondegenerate setting if $e, e'$ are two errors such that $\sigma(e) = \sigma(e')$ and $p_E(e) > p_E(e')$ then $(\sigma(e'), e')$ is a bad example. The total probability of bad examples is $\Pr(\text{bad}):=\Pr_{(\sigma, y)}(y\neq f^*(\sigma))$.
    \item An \textit{important example} is any good example $(\sigma, y)$ that the baseline decodes incorrectly, $f_0(\sigma) \neq y$. The importance of an example is always defined with respect to some $f_0$. The total probability of important examples is $\Pr(\text{important}):=\Pr_{(\sigma, y)}(y= f^*(\sigma), y \neq f_0(\sigma))$.  An \textit{unimportant} example is a good example that is not important. 
\end{itemize}

In general, $\Pr(\text{good}) \geq \Pr(\text{bad})$, but there are usually more possible bad examples than good examples. Bad examples appear because (i) every $[[n, k]]$ quantum code associates each syndrome $\sigma$ with one of $2^{n+k}$ physical errors (nondegenerate) or one of $2^{2k}$ logical errors (degenerate), meaning that there are $2^{n+k} - 1$ or $2^{2k}-1$ bad examples respectively (ii) bad examples can result from noisy syndromes, which we do not consider here. Each dataset $D_N$ contains mostly unimportant examples that a baseline decoder is already capable of decoding, mixed in with bad examples that confuse the neural decoder by providing an incorrect decoding example. Intuitively, bad examples act like noise in the training process -- a comparison that we formalize in Proposition~\ref{prop:1}. 

The challenge of $\DDD$ is that important examples are usually rare and occur at similar frequencies to bad examples. Often, a baseline decoder can correctly decode errors up to some weight $t$, so that $\Pr(\text{important})$ is at most the total probability of errors with $\wt(e)\geq t+1$. If a device has an error rate of $p$ per (qu)bit, then $\Pr(\text{important})$ can shrink exponentially with $t$.  We can quantify the rarity of important examples by defining the \textit{importance} of an example with respect to a baseline decoder $f_0$ and an $\MLD$ $f^*$ as 
\begin{equation}\label{eq:importance}
    J((\sigma,y); f_0) := p_{\Sigma Y}(\sigma, y) \I\{f^*(\sigma) = y\} \I \{ f_0(\sigma) \neq y\},
\end{equation}
where $\I$ denotes the indicator function. Eq.~\ref{eq:importance} relates example importance to learning:  the cumulative importance of examples in a dataset is an upper bound on the improvement of any neural decoder $\hat{f}$ trained according to $\DDD$ over the baseline decoder:
\begin{align}\label{eq:imp_ub}
    \Pr_{(\sigma,y)\sim \mD} &(\hat{f}(\sigma) = y) - \Pr_{(\sigma,y)\sim \mD} (f_0(\sigma) = y)  \leq \sum_{(\sigma,y)} J((\sigma,y); f_0) := \Pr(\text{important}).
\end{align}
The gap in this inequality is due to the difference in the accuracy of any $\hat{f}$ vs. $f^*$ as well as the probability that $f_0$ correctly decodes a bad example, which tends to be small (see Appendix~\ref{app:importance}). The cumulative importance of a distribution of examples directly describes the (best) relative performance of a neural decoder: In the extreme case where a distribution contains no important examples, a neural decoder cannot learn to decode with accuracy any better than the baseline $f_0$. A limitation of example importance is that evaluating $J$ requires access to an $\MLD$, and may be intractable due to the hardness of optimal decoding in general. More broadly, this implies that determining whether a dataset is useful for improving a neural decoder is an intractable problem. 

\subsection{Learning to decode the repetition code}\label{sec:class_repe}

We will first explore $\DDD$ of a simple toy problem: Decoding the (classical) $n$-bit repetition code with a biased bitflip error model. The $\MLD$ for this code is trivial given knowledge of the error model or a small dataset of examples, so what can be learned by attempting to solve $\DDD$ on such a simple problem? We want to understand how well basic neural networks can decode QECC syndromes, but training a single neural decoder on even modestly-sized tasks like a multi-round surface code experiment is stochastic and expensive. By investigating a toy problem that is theoretically easy to decode, we can efficiently and systematically study different architectures for neural decoders and show that they struggle with this trivial instance of $\DDD$ without special intervention. This will establish the relevance of example importance and the need for specialized training techniques in the more difficult setting of decoding QECCs. Our first result for studying data-driven decoding relates $\DDD$ for this toy problem to a noisy classification problem:

\begin{proposition}[$\DDD$ \textbf{for the repetition code is noisy binary classification, given $f_0$}]\label{prop:1}
Let $D_N =\{(\sigma_i, e_i)\}_{i=1}^N$ be  a dataset sampled according to $\DDD$ for the repetition code, i.e. $\sigma_i = He_i$ for the $n$-bit parity check matrix $H$. Fix a baseline decoder $f_0$, and define a labeling scheme $z(\sigma) = \I\{f_0(\sigma) = f^*(\sigma)\}$. Then, there is noisy labeling scheme $\tilde{z}(\sigma)$ such that solving $\DDD$ is equivalent to finding $\hat{f}: \{0,1\}^{n-1}\rightarrow \{0,1\}$ that maximizes the binary classification accuracy $\Pr_{\sigma, z\sim p_{\Sigma Z}}(\hat{f}(\sigma) = z)$ given the noisy dataset
\begin{equation}
    \tilde{D}_N := \{(\sigma_1, \tilde{z}_1),\dots, (\sigma_N, \tilde{z}_N)\}.
\end{equation}
Furthermore, the overall probability of label noise for both classes equal to $\Pr_\sigma (\tilde{z} \neq z) = \Pr(\text{bad})$, and for $\xi_n = \exp\left(-\frac{n}{2} (1-2p)^2\right)$ the class prior probabilities obey
\begin{align}
    |\Pr_{\sigma} (z = 0) - \Pr(\text{important})|   &\leq  \xi_n , \label{eq:prop1.1} \\
    |\Pr_{\sigma} (z = 1) - \Pr(\text{unimportant})|  &\leq  \xi_n \label{eq:prop1.2}
\end{align}
\end{proposition}
Proposition~\ref{prop:1} (proof in Appendix~\ref{app:toymodel}) formalizes the amount of important examples and bad examples as measures of class imbalance and label noise, respectively. Eq.~\ref{eq:prop1.1} shows that the probability of bad examples is related to an overall noise probability, and Eqs.~\ref{eq:prop1.1}-\ref{eq:prop1.2} show that for small $np$, the difference in the frequency of important examples vs. unimportant examples is directly linked to an imbalanced classification problem. We use this result to directly characterize how the presence of bad examples affects solutions for $\DDD$:

\begin{corollary}[\textbf{Sample complexity of $\DDD$ for the repetition code, given $f_0$}]\label{cor:1}
    Let $\mathcal{F}$ be the set of boolean functions on $n-1$ bits. For a fixed instance of $\DDD$ and baseline decoder $f_0: \{0,1\}^{n-1} \rightarrow \{0,1\}^n$ with logical error probability $\epsilon := \Pr_{\sigma, e}(f_0(\sigma) \neq e)$, then with probability at least $1- \delta$, a neural decoder can learn a better decoding strategy than $f_0$ given a dataset of size
    \begin{equation}
        N \leq \frac{2}{(1-2\Pr(\text{bad}))^2 \epsilon^2 }\log \left(\frac{|\mathcal{F}|}{\delta}\right). 
    \end{equation}
\end{corollary}
While this bound grows impractically loose for large $n$, it captures how the \textit{sample complexity} of $\DDD$ grows monotonically with $\Pr(\text{bad})$ (proof in Appendix~\ref{app:cor2}). In Appendix~\ref{app:class_imbalance} we discuss further how the noise and class imbalance described in Proposition~\ref{prop:1} and Corollary~\ref{cor:1} (which describe an equivalent classification problem) manifest in the $\DDD$ setting under reasonable assumptions.

In light of Proposition~\ref{prop:1}, data augmentation techniques for providing additional decoding examples in $\DDD$ can introduce a tension: A large class imbalance (small $\Pr(\text{important})$) makes learning difficult, and so a large body of work \cite{japkowicz2002class,heLearningImbalancedData2009,krawczyk2016learning} suggests that data augmentation might improve neural decoder performance. However, these techniques can introduce additional label noise by also increasing the frequency of bad examples (Fig.~\ref{fig:rep_code_importance}a), thereby increasing the sample complexity of $\DDD$ by Corollary~\ref{cor:1}. This suggests a tradeoff between having more important examples (leading to a more balanced classification problem) and introducing more bad examples (leading to a noisier classification problem), without any clear resolution. So, we will study this tradeoff numerically.

\subsection{Experiments for the repetition code}

Having developed a theory of $\DDD$ for the classical repetition code, we now demonstrate some implications of this theory for training neural decoders. We focus on $\DDD$ for the $n=8$, distance $d=8$ repetition code with biased noise because (i) even large neural decoders struggle with this task given a reasonable amount of training data (ii) by using small $n$ we can efficiently train thousands of neural decoders to demonstrate patterns in training behavior (iii) we can thoroughly study \textit{turning the knob} -- a simple data augmentation technique -- and empirically observe the resulting tradeoff between noise and class imbalance discussed in the previous section. 

To demonstrate the challenges of data-driven decoding in a straightforward way, we can consider a simple error model where half of the bits flip with probability $p < 1/2$ while the other half flip with probability $\alpha p$ with $\alpha < 1$. This error distribution, denoted as $p_E[p, \alpha]$, is inspired by real devices where error rates may not be identical for different bits. For our baseline we consider the standard \textit{minimum weight decoder}  $f_0(\sigma) \in \argmin_{e: He=\sigma}\wt(e)$ with ties broken randomly. For an iid bitflip error model, $f_0$ is a maximum likelihood decoder. For a biased error model, $f_0$ still decodes most syndromes correctly but sometimes fails to decode examples for which the weight of the error is close to $n/2$ (see Fig.~\ref{fig:rep_code_importance}a). The challenge for a neural decoder in the toy problem is to beat $f_0$ by correctly decoding these important errors with weight close to $n/2$.

However, learning from important examples is challenging for neural decoders: If we let $\bar{p} = (p + \alpha p)/2$, then sampling pairs $(\sigma, e)$ according to $p_E[p,\alpha]$ usually yields errors with weights near $n\bar{p} \ll n/2$. The probability of sampling a weight $n/2$ error is suppressed exponentially, e.g. $\Pr(\wt(E) = n/2) \lessapprox (4\bar{p})^{n/2}$ \cite{2797618}. This is extremely unfavorable scaling for our experiments where $p=0.1, \alpha=0.7$ on $n=8$ bits, as only about $1/400$ of examples in $D_N$ will be important. Yet, these are precisely the examples that a neural decoder needs in order to outperform $f_0$. Meanwhile, an $\MLD$ for this problem can (theoretically) be learned from few data since the parameter $\alpha$ is learnable with $>99\%$ probability given $N=(p - \alpha p)^{-2} \approx 2000$ examples. Providing even $N=2000$ examples is generous since in real experiments we will rather expect $N \ll 2^n$. To account for the possibility of memorization, we compare neural decoders' LEP to a lookup table for $D_N$ defined as $f_{D_N}(\sigma) \in \argmax_{e} \hat{p}_{\Sigma E}(\sigma, e)$, where $\hat{p}_{\Sigma E}$ is a histogram of $(\sigma, e)$ pairs in $D_N$.

\subsection{Importance sampling by turning the knob} 

The lack of important examples in the training dataset $D_N:= D_N[p, \alpha]$ suggests that we might improve the accuracy of a neural decoder by doing training on some \textit{other} dataset $D_N[\beta p, \alpha]$ generated using a higher error rate $\beta p>p$. By increasing $\beta$ we are essentially \textit{turning the knob} that controls the underlying error rate of a training dataset. This technique was first introduced in Ref.~\cite{chamberland_deep_2018}, though has been used sparingly since then (see Sec.~\ref{sec:prior} for a review of prior work). In contrast to prior works, we analyze this technique systematically and motivate its performance through the theory of example importance. 

Turning the knob is interesting because it can, in principle, be implemented on real hardware. This not only increases the number of ordinarily-unlikely errors in the training data, but ideally generates these data according to a distribution that is \textit{qualitatively similar} to the true, underlying error behavior of the device. This technique is firmly grounded in the theory of importance sampling: Correctly turning the knob will increase the fraction of important examples in the training data without introducing too many additional bad examples.

We first simulate this technique in the $\DDD$ setting for the repetition code by training neural decoders on a dataset $D_N[\beta p, \alpha]$ sampled from $p_E[\beta p, \alpha]$ rather than the original training data $D_N[p, \alpha]$. By Proposition~\ref{prop:1} this results in a more balanced -- but noisier -- classification problem. Fig.~\ref{fig:rep_code_importance}b shows that turning the knob at initially robustly improves the performance of neural decoders, across a variety of architectures (FNNs, CNNs, transformer), hyperparameters, and randomness in datasets and initializations (see App.~\ref{sec:experiments} for details). In all cases, the best model trained on $D_N[\beta p, \alpha]$ with $\beta \gg 1$ achieved $5\times$ to $10\times$ lower LEP than the best model trained on $D_N[p, \alpha]$, while still generalizing to unseen test examples.

Turning the knob starts to backfire when $\beta \gtrapprox 3$, resulting in a characteristic ``U''-shaped curve in LEP versus $\beta$. We attribute this to two effects: First, Corollary~\ref{cor:1} suggests that as $\Pr(\text{bad})$ increases more samples are needed for learning. Second, increasing $\beta$ can cause sudden changes in signal-to-noise ratio: for a large enough $\beta$, the best decoder for $D_N[\beta p, \alpha]$ is no longer the best decoder for $D_N[p, \alpha]$, implying that some important example for $p_E[\beta p, \alpha]$ is a bad example for $p_E[ p, \alpha]$. We show that such an example appears once $\beta$ is large enough to satisfy 
\begin{equation}\label{eq:threshold}
     \frac{\alpha \beta p)^2}{(1 - \alpha \beta p )^2} < \frac{(\beta p)^4}{(1 - \beta p)^{4}}
\end{equation}
(see App.~\ref{sec:well_ordered}). For $p=0.1, \alpha=0.7$ this gives $\beta \approx 3.7$, which coincides with a decline in neural decoder performance in  Fig.~\ref{fig:rep_code_importance}b and the resulting ``U''-shaped curve.

\section{Importance in QECCs: Beyond the toy problem}

We now analyze the performance of neural network decoders for $\DDD$ on quantum error-correcting codes. Our experiments are be similar to those in the previous section: We study turning the knob using small QECCs, which allows us to test the technique's robustness in the presence of large stochastic effects that appear during neural network training. Using small QECCs will also allow us to study how the $\MLD$ for each code begins to fail as a function of knob factor $\beta$. Neither of these analyses is possible for large QECCs, motivating our study of codes with $n\leq 9$ qubits.

There are important qualitative differences in $\DDD$ data for QECCs versus classical codes. In the classical repetition code, important examples correspond to errors with weight $d/2$, and example importance diminishes rapidly with weight. But for an $[[n, k, d]]$ QECC it is not always true that errors with weight just above $d/2$ have the highest importance. The $[[5, 1, 3]]$ code \cite{5qubit1,5qubit2} is a counterexample, since the $\MLD$ for this code can never correct two-qubit errors\footnote{Every $e \in\mS$ has $\wt(e)=4$, and so for any error $t$ with $\wt(t)=1$ the maximum likelihood coset $t\mS$ can only contain errors with weight 3, 4, or 5.}. Furthermore, it is harder to formally characterize $\DDD$ in QECCs: While Proposition~\ref{prop:1} showed that $\DDD$ for a $k=1$ classical decoding problem is equivalent to a noisy, binary classification problem, degenerate $\DDD$ for an $[[n, k]]$ QECC is a multiclass classification problem that requires predicting one of $4^k$ logical errors given each syndrome. For two examples $(\sigma, y)$ and $(\sigma, y')$ it might be true that $y\neq y'$ but $y,y'$ are both maximum likelihood. Since a neural decoder has deterministic outputs, only one of $y$ or $y'$ can be important and the other one must be bad. While a rigorous characterization of degenerate $\DDD$ of QECCs is out of reach for now, the conceptual relationship between $\Pr(\text{bad})$, $\Pr(\text{important})$, noise rates, and class imbalance still provides useful intuition for $\DDD$ experiments involving QECCs.

\subsection{\texorpdfstring{$\DDD$}{DDD} in quantum codes}\label{sec:qDDD}

\begin{figure}[ht]
    \centering
    \includegraphics[width=\linewidth]{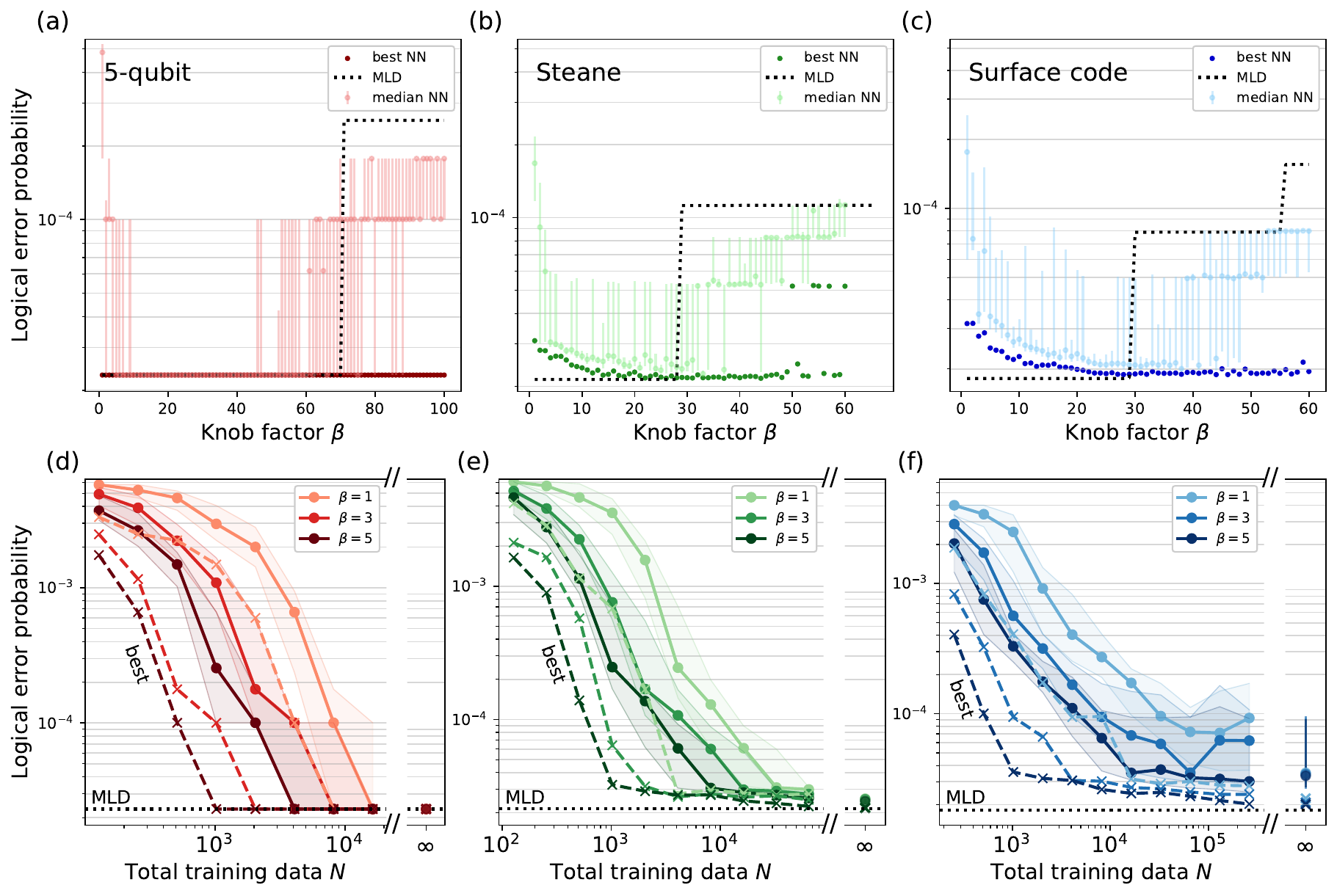}
    \caption{ \textbf{(a-c) Importance sampling (via turning the knob) reliably and robustly improves decoder error rates for small codes.} Each point depicts median LEP of $\sim 60$ transformer decoders trained, with bars showing IQ range with respect random initializations and hyperparameter range (see Appendix~\ref{app:dddexp}), and a solid line shows the best LEP. In all codes, the median LEP begins to rise when $\beta$ is large enough that the $\MLD$ for $p_E[\beta p]$ is subpotimal for $p$, suggesting a strategy for finding a good knob factor $\beta$ (see Sec.~\ref{sec:opt_beta}). \textbf{(d-f) Training via turning the knob can reach similar LEP using $\sim 1/10$ as many training data.} In the limit of infinitely many training data (with gradient updates computed directly from $p_E[\beta p]$), training runs with different $\beta$ converge towards a similzar LEP, as a scarcity of important examples no longer hinders trainability of each neural decoder. ``$\times$'' shows the best of $\sim 150$ transformer-based decoders, $\bullet$ shows median, shading shows IQ range.}
    \label{fig:toric_results_new}
\end{figure}

We validate importance sampling for QECCs by showing that turning the knob remains an effective strategy for improving neural decoders for several (small) codes. We study $\DDD$ for the $[[5, 1, 3]]$ 5-qubit code \cite{5qubit1,5qubit2}, the $[[7, 1, 3]]$ Steane color code \cite{steane1996multiple}, and the $[[9, 1, 3]]$ rotated surface code \cite{bravyi1998quantumcodeslatticeboundary,freedman1998projectiveplaneplanarquantum}, each with a single round of perfect syndrome measurement. We consider an error model where errors are independent but non-identical across qubits: each qubit is affected by depolarizing errors with a rate $p_i$, where the set $\{p_i\}_{i=1}^n$ is sampled independently with a mean of $p$ and variance $\sigma^2$; we write the resulting probability distribution over $\mPhat_n$ as $p_E[p]$. 

Figs.~\ref{fig:toric_results_new}a-c show the effect of turning the knob on each code with true error rates $p=10^{-3}$ with $\sigma^2 = 10^{-3}$. We turn the knob by training neural decoders for $\DDD$ using a dataset $D_N[\beta p]$ sampled by multiplying each independent depolarizing error probability by $\beta$ (so that the set of error probabilities is $\{\beta p_i\}$). In all cases, turning the knob improves median neural decoder LEP with respect to a pre-determined \textit{good} range of hyperparameters (tuned such that 70\% of models achieve $\MLD$ when trained on only good examples --see Appendix~\ref{app:dddexp}). More importantly, the \textit{best} decoder LEP improves with increasing $\beta$ for the Steane and Surface codes. 

As was the case with $\DDD$ for the classical repetition code, the logical error probability (LEP) of neural decoders trained on $D_N[\beta p]$ exhibits a ``U'' shape with respect to the knob factor $\beta$. While the LEP initially decreases with $\beta$, once $\beta$ becomes large there is some syndrome $\sigma$ such that the most likely logical error given $\sigma$ with respect to $p_E[\beta p, \sigma^2]$ differs from the $\MLD$ logical error with respect to $p_E$. This pattern depends on the training set $D_N[\beta p]$ encoding prior knowledge of the error model. In Appendix~\ref{app:toric_extra} we repeat some experiments for the $[[9, 1, 3]]$ code by training neural decoders using an error model with independent and identical error rates $p_i=\beta p$, and find that the technique largely fails.

When neural decoders are trained to minimize decoding error on the higher error rate training data $D_N[\beta p]$, they learn the decoding rule that approaches the $\MLD$ for $p_E[\beta p]$. Thus, when $\beta$ is large enough that the $\MLD$ for $p_E[\beta p]$ begins to fail at decoding errors from $p_E[p]$, the performance of the neural decoder must also begin to degrade. We will now formalize this observation, and use this behavior to propose a technique to search for a good knob factor

\subsubsection{Searching for a good knob factor}\label{sec:opt_beta}

A challenge of importance sampling for data-driven decoding is that the neural decoder trained using higher error rates may become inaccurate on the original distribution of syndromes. This can be understood in terms of the disagreement between maximum likelihood decoders on the two different datasets. Let $p$ denote the true distribution of syndromes for a given error model (with $\MLD$ denoted as $f^*$), and let $p_\beta$ denote the distribution of errors after turning the knob (with corresponding $\MLD$ $f^*_\beta$). Then, any decoder $\hat{f}$ obeys
\begin{equation}
        \Pr_{\sigma \sim p}\left(\hat{f}(\sigma) \neq f^*(\sigma)\right)  \leq \Pr_{\sigma \sim p_\beta}\left(  \hat{f}(\sigma) \neq f_\beta^*(\sigma) \right) + 2 d_{TV}(p, p_\beta )  + \underbrace{\Pr_{\sigma \sim p}(f^*(\sigma) \neq f_\beta^*(\sigma) )}_{\text{misalignment}} , \label{eq:surrogate_ub}
\end{equation}
where $d_{TV}(p,q):=\frac{1}{2}\sum_{\sigma}|p(\sigma) - q(\sigma)|$ denotes the \textit{total variation distance} of two distributions. The goal in $\DDD$ is to minimize the LHS. If $\hat{f}$ is trained using syndromes from $p_\beta$, then the first term on the RHS decreases with training accuracy, while the second term increases gradually with increasing $\beta$. However, the \textit{misalignment} of these distributions -- the LEP of an $\MLD$ for $p_\beta$ evaluated on $p$ -- can increase sharply as an $\MLD$ for $p_\beta$ becomes suboptimal for decoding syndromes from $p$. 

When a neural decoder trained on syndrome data sampled from $p_\beta$ (i.e. importance sampling) is evaluated on syndrome data from the true error distribution $p$, then the neural decoder's LEP can become large whenever $p$ and $p_\beta$ are misaligned in the above sense. This suggests a heuristic for turning the knob: Continue to increase $\beta$ until $p$ and $p_\beta$ become misaligned. Figs.~\ref{fig:toric_results_new}(a-c) show that misalignment between $p_\beta$ and $p$ (dotted black line) roughly coincides with the $\beta$ minimizing median decoder error, validating this heuristic.

\subsection{Sample importance with multiple rounds of decoding}\label{sec:detectorddd}

One limitation of $\DDD$ is that it provides labeled data that are generally impossible to sample from a real quantum device due to the non-clonability of quantum information. So we would now like to study an experimentally-realizable extension of $\DDD$. For a CSS code with stabilizer generators $\{x_1, \dots, x_{r_X}, z_1, \dots, z_{r_Z}\}$, consider preparing an eigenstate of $\bar{Z} \in \mL$, followed by $T$ consecutive rounds of stabilizer measurements and a final measurement in the basis of $\bar{Z}$. The result is a labeled tuple of data $(V_X, V_Z,y)$ , where $V_X = \{\langle x_i\rangle^{(t=1)} \dots, \langle x_i\rangle^{(t=T)} \}_{i=1}^{r_X}$, $V_Z = \{\langle z_i\rangle^{(t=0)}, \dots, \langle z_i\rangle^{(t=T+1)} \}_{i=1}^{r_Z}$, and $y = \I\{ \langle \bar{Z}\rangle^{(t=T+1)} \neq \langle \bar{Z}\rangle^{(t=0)}\}$ i.e. $y$ is 1 if the observable $\bar{Z}$ has flipped at timestep $T$. Repeat this procedure $N$ times to create a dataset $ D_N:=\{(V_X^i, V_Z^i, y_i)\}_{i=1}^N$. The learner is given $D_N$ and tasked with predicting $y$ (i.e. a logical Pauli-$Z$ error) given $(V_X, V_Z)$. This problem, which we will call $\DDDD$, captures the task of data-driven decoding  with multiple measurement rounds studied in Refs.~\cite{chamberland_deep_2018,baireuther_machine_2018,varsamopoulos2019decodingsurfacecodedistributed,
zhang_scalable_2023,varbanov_neural_2023,chamberlandTechniquesCombiningFast2023,lange_data_driven_2023,bausch_learning_2023,PhysRevResearch.6.L032004}. In contrast with the single-round $\DDD$ problem that involves training data that cannot be sampled from a real device, $\DDDD$ describes a learning problem involving data that can be experimentally measured from a series of QECC experiments.

As with $\DDD$, we can reframe $\DDDD$ as an imbalanced binary classification problem. However, examples $(\sigma, y)$ that are important with respect to the $\DDDD$ problem are not necessarily important with respect to decoding a given syndrome, since the $\DDDD$ data only captures the $\bar{X}$- or $\bar{Z}$-component of a logical error. Instead, we will consider optimal and baseline \textit{classifiers} rather than decoders. Let $\mathcal{V}$ denote the space of possible $T$-round stabilizer measurement outcomes in $\DDDD$. We define $y^*: \mathcal{V} \rightarrow \{0,1\}$ to be an optimal classifier  $y^*(\sigma):= \argmax_{y\in\{0,1\}}\Pr(Y=y|\Sigma=\sigma)$, and consider any baseline $y_0:  \mathcal{V} \rightarrow \{0,1\}$. We now define important and unimportant examples in the $\DDDD$ problem as follows:
\begin{align}
    \Pr(\text{unimportant}) &:= \Pr_{(\sigma, y)}(y^*(\sigma)=y, y_0(\sigma)= y) \\
    \Pr(\text{important}) &:= \Pr_{(\sigma, y)}(y^*(\sigma)=y, y_0(\sigma)\neq y)
\end{align}
We can immediately characterize $\DDDD$ as an imbalanced classification problem:
\begin{proposition}[$\DDDD$ \textbf{is equivalent to noisy binary classification, given $y_0$}]\label{prop:2}
    Consider the $\DDDD$ problem for any $[[n, 1, d]]$ QECC under the effect of iid depolarizing noise model with probability $p$ of error. Then, letting $\xi_n = \exp\left(-(d - 2np)^2/2n\right)$, $\DDDD$ is equivalent to a noisy, imbalanced binary classification problem for labels $z \in \{0,1\}$ having class priors
\begin{align}
    |\Pr_{\sigma}(z(\sigma) = 0) - \Pr(\text{important})| &\leq \xi_n, \\ 
    |\Pr_{\sigma}(z(\sigma) = 1) - \Pr(\text{unimportant})| &\leq \xi_n.
\end{align}
\end{proposition}
Any high-accuracy baseline decoder $f_0$ can be used as a high-accuracy baseline classifier $y_0$, and so it remains true that $\Pr(\text{important})\ll \Pr(\text{unimportant})$ in most realistic settings. Therefore, by analogy with the $\DDD$ problem, Proposition~\ref{prop:2} suggests that (i) turning the knob should remain an effective strategy for learning to decode for realistic and complex decoding problems and (ii) increasing training data error rates should result in a ``U''-shaped curve for the logical error probability.

\begin{figure}
    \centering
    \includegraphics[width=\linewidth]{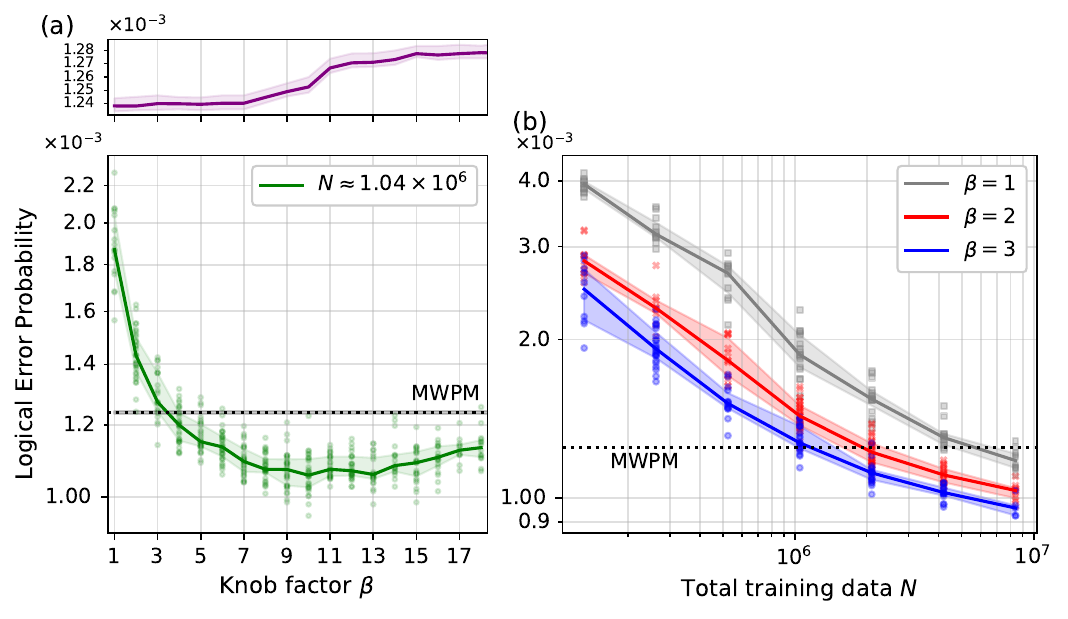}
    \caption{\label{fig:gnn_results_val}\textbf{Turning the knob reliably improves neural decoders in $\DDDD$ (multiple rounds of decoding).} \textbf{(a)} Gneural decoders \cite{lange_data_driven_2023} trained to decode the $d=3,T=5$ rotated surface code exhibit the same ``U''-shaped curve when turning the knob, across initializations and hyperparameters for $\beta \in \{1, \dots, 19\}$ ($10\text{-}30$ GNNs per $\beta$, $435$ total). Top plot shows LEP of MWPM on a matching graph with edge weights corresponding to noise rate $\beta p$. \textbf{(b)} Turning the knob is advantageous for all dataset sizes in $N\in\{2^{17}, \dots, 2^{23}\}$. The improvement persists even after controlling for the difference in the number of trivial examples in $D_N[\beta p]$  vs. $D_N[p]$ (see Appendix~\ref{app:additional}), suggesting that sampling $p_E[\beta p]$ yields a qualitatively superior training distribution. All plots: solid line is median test error, shading is IQ range of provided points.}
\end{figure}

We consider a $[[9, 1, 3]]$ rotated surface code with $T=5$ rounds of syndrome measurement and circuit-level errors at a base rate of $p=10^{-3}$. We denote the training dataset and error distribution corresponding to this error model as $D_N[p]$ and $p_E[p]$ respectively. We trained Graph Neural Network (GNN) decoders using the architecture and methods of Ref.~\cite{lange_data_driven_2023} on datasets $D_N[\beta p]$ with cross-validation with respect to $D_N[p]$.  Fig.~\ref{fig:gnn_results_val}b shows similar performance improvements across a wide range of training dataset sizes $N$. This robustly reproduces anecdotal observations from Refs.~\cite{chamberland_deep_2018,varbanov_neural_2023}, but conflicts with observations made in Ref.~\cite{chamberlandTechniquesCombiningFast2023} (though our training scheme differs by using cross-validation on $\beta=1$ data). See Appendix~\ref{sec:DDDD_exp} for more details on the model architecture, training procedure, and error model. 

Turning the knob predictably begins to fail when the $\MLD$ for $p_\beta$ becomes misaligned, and therefore less capable of decoding errors sampled from $p$ (see Sec.~\ref{sec:opt_beta}). In practice, this behavior cannot be used to predict a good $\beta$ since evaluating either $\MLD$ is intractable. However, we can evaluate the misalignment between a data-driven baseline calibrated for $p_\beta$ versus one calibrated for $p$ (such baselines may even be data-driven, e.g. Ref.~\cite{sivak2024optimizationdecoderpriorsaccurate}). For a baseline $h_q$ calibrated on errors with distribution $q$, we can use misalignment in $h_q$ as a surrogate for misalignment in the $\MLD$: The heuristic for a good knob factor is the largest $\beta$ such that $\Pr\left(h_{p_{\beta\theta}}(\Sigma) \neq h_{p_\theta}(\Sigma)\right) \leq \delta$ for some error parameter $\delta$. 

This heuristic is validated in Fig.~\ref{fig:gnn_results_val}a: MWPM using a matching graph weighted according to $p_E[\beta p]$ drops in accuracy near the $\beta$ that is empirically optimal for training neural decoders. Notably, $\beta$ values smaller than this still result in substantial improvements in neural decoder LEP compared to $\beta=1$.

\section{Related work}\label{sec:prior}

\textbf{Data-driven decoding} has been demonstrated to outperform standard decoders for classical  LDPC codes in the presence of both simple and more exotic noise channels (e.g. Refs.~\cite{KimJRKOV18,8815400,nachmani2019hyper,pmlr-v130-garcia-satorras21a,choukroun_error_2022,cammerer2022graph,kwak2023boosting}), as well as experimental QECCs \cite{bausch_learning_2023}. Data-driven decoders have been studied for QECCs in the nondegenerate $\DDD$ setting~\cite{torlai_neural_2017,krastanov_deep_2017,varsamopoulos_decoding_2018,varsamopoulos_comparing_2020,breuckmann_scalable_2018,davaasuren_general_2020}, degenerate $\DDD$ setting~\cite{varsamopoulos2019decodingsurfacecodedistributed,maskara_advantages_2019,davaasuren_general_2020,wagner_symmetries_2020,ni_neural_2020, bhoumikEfficientDecodingSurface2021,overwater_neural_network_2022,meinerz_scalable_2022,cao_qecgpt_2023,gicev_scalable_2023,wang_transformer_qec_2023,egorovENDEquivariantNeural2023,Choukroun_Wolf_2024}, and $\DDDD$ setting~\cite{chamberland_deep_2018,baireuther_machine_2018,varsamopoulos2019decodingsurfacecodedistributed,
zhang_scalable_2023,varbanov_neural_2023,chamberlandTechniquesCombiningFast2023,lange_data_driven_2023,bausch_learning_2023,PhysRevResearch.6.L032004}. Our work specifically addresses training such decoders, and we have not considered other important experimental constraints such as throughput or realistic data. Our formalism is less relevant for graph-based decoding algorithms \cite{liu_neural_2019,ninkovic2024decodingquantumldpccodes,maan2024machinelearningmessagepassingscalable} and reinforcement learning approaches \cite{sweke_reinforcement_2021,matekole_decoding_2022,Andreasson_2019,PhysRevResearch.2.023230}. 

\textbf{Importance sampling for data-driven decoding} has been used in previous work to improve the accuracy of neural network decoders. Ref.~\cite{chamberland_deep_2018} first showed that training with $\beta>1$ (turning the knob) could improve NN decoder accuracy. The technique was later applied to QECCs with limited details~\cite{liu_neural_2019,varbanov_neural_2023}. Ref.~\cite{chamberlandTechniquesCombiningFast2023} observed that turning the knob could hurt performance. In contrast with prior works, we provide theory to justify this technique and demonstrate robust improvement with respect to hyperparameter choices, randomness in data sampling and weight initialization, NN decoder architecture, and choice of codes. Some techniques have been developed specifically for importance sampling in QEC simulations \cite{Bravyi_2013,mayer2025rareeventsimulationquantum}, though these cannot clearly be extended to be implemented on hardware. 

\textbf{Curriculum learning} \cite{bengio2009curriculum} involves initially training a model only on easy examples, and then later providing difficult examples. Common measures of example difficulty are (i) complexity of the learning task, (ii) the diversity of the input data, and (iii) the amount of (label) noise \cite{wang2022}. Proposition~\ref{prop:1} shows that \textit{turning the knob} is related to curriculum learning in the sense of (ii)-(iii), as suggested by Ref.~\cite{bausch_learning_2023}: a model trained on $D_N$ with small $\beta$ sees fewer distinct training examples and fewer noisy examples. Prior works in data-driven decoding have done curriculum learning by generating training data with $\beta \leq 1$ \cite{maskara_advantages_2019,bausch_learning_2023}. However, curriculum learning usually involves downsampling to decrease noise/complexity \cite{sovianyCurriculumLearningSurvey2022}, whereas turning the knob involves data augmentation that \textit{increases} noise rates while decreasing class imbalance, leading to a characteristic ``U''-shape in Figs.~\ref{fig:rep_code_importance}-~\ref{fig:toric_results} due to a higher proportion of bad training examples. 

\textbf{Learning with class imbalance and noise} (e.g. Refs.~\cite{japkowicz2002class,heLearningImbalancedData2009,krawczyk2016learning} and \cite{frenay2013classification,xiao2015learning,song2022learningnoisylabelsdeep}) are widely studied settings in machine learning. Proposition~\ref{prop:1} explicitly connects $\DDD$ to these settings for a simple decoding problem. While this connection is less explicit for degenerate decoding with $k>1$, a core characteristic of $\DDD$ and $\DDDD$ is that outperforming existing baselines requires learning from a long tail of noisy, low-probability examples.

\section{Discussion}

\paragraph{Limitations} The main shortcoming of example importance is that identifying important examples can be intractable in large system sizes. Thus, our theory can only provide sharp characterizations for small instances of systems, with the hope of either (i) providing intuition that extends to larger systems or (ii) motivating heuristics for estimating importance, for example approximating $\MLD$ ``offline'' to prepare a dataset for pretraining a neural decoder before fine-tuning on experimental data. 

Similarly, analytically determining the optimal knob factor $\beta$ to achieve the  minimum of the ``U'' shape in Figs.~\ref{fig:rep_code_importance}-\ref{fig:toric_results} will be intractable, though our results suggest this optimum could be determined empirically with sufficiently many experiments. To achieve the best performance, turning the knob requires partial knowledge of $p_E$ to simulate a \textit{qualitatively similar} distribution with higher error rates, which might be experimentally infeasible -- though we note that neural decoders have been successfully pre-trained on simulated data using precisely this technique by incorporating device calibration data into simulations \cite{varbanov_neural_2023}. Furthermore, Fig.~\ref{fig:toric_results}c shows that prior knowledge of $p_E$ may not be necessary for improving accuracy via turning the knob given limited training data from $p_E$, but it remains unclear under what conditions this might hold. Finally, we have mostly ignored the throughput of decoders, which is an essential parameter for experimentally viable decoders. 

\paragraph{Future work} Our work leaves several questions open regarding importance sampling for data-driven decoding

\begin{itemize}
    \item \textbf{An experimentally realizable protocol?} Our results motivate the development of an \textit{experimentally realizable protocol} for turning the knob. As we have shown, this requires sampling training data with higher error rates, but from a distribution that is \textit{qualitatively similar} to the ordinary error behavior of the device. One possibility is to experimentally sample a $\DDDD$ dataset after intentionally increasing error rates on the experimental device, for example by idling longer between each gate operation or before each set of syndrome measurements. Related strategies have been successfully applied for quantum error mitigation, e.g. ``zero-noise extrapolation'' \cite{li2017efficient,temme2017error,kandala2019error}. Furthermore, we have seen that turning the knob is primarily effective on devices for which the error rate is already small.
    \item \textbf{When do we turn the knob?} We have analyzed how neural decoder errors follow a U-shaped curve with respect to the knob factor $\beta$ as suggested by Propositions\ref{prop:1} and \ref{prop:2}. However, there is some interplay between $\beta$ and the number of training data $N$. Figs \ref{fig:toric_results_new}(d-f) show that turning the knob is not substantially better than ordinary training for sufficiently large datasets. More generally, the effectiveness of turning the knob may depend on the dataset size in complex ways.
    \item \textbf{Can we predict the optimal $\beta$?} Both theory and numerics in this work suggest the existence of an optimum knob factor $\beta^*$. Propositions~\ref{prop:1} and \ref{prop:2} show that the tradeoff leading to $\beta^*$  is intrinsic to an error model rather than a particular decoder. We have described several heuristics for guessing $\beta^*$, though a provably correct scheme for optimizing $\beta$ remains out of reach.
    \item \textbf{Can we increase sample importance in other ways?} We have focused on turning the knob since it is a simple data augmentation technique with some prior interest in the literature. However, example importance is broadly applicable to any data augmentation technique in $\DDD$ and $\DDDD$. While existing techniques can be used for importance sampling in simulations  \cite{Bravyi_2013,mayer2025rareeventsimulationquantum}, it remains to be seen whether the data generated by these techniques can effectively capture the idiosyncratic error behavior of any particular quantum device.
\end{itemize}

\section{Conclusion}

We have presented a theory of example importance that characterizes the usefulness of training data for learning to decode (quantum) error-correcting codes. We used this theory to precisely explain the peculiarities of data-driven decoding ($\DDD$), showing that a simple instance of $\DDD$ is equivalent to noisy binary classification with class imbalance. This theory suggests that augmenting training data by \textit{turning the knob} to generate decoding examples using higher device error rates leads to a tradeoff between higher label noise rates and lower class imbalance. We characterized this tradeoff empirically using thousands of learning experiments involving different learning architectures and codes, demonstrating a regime of robust improvements to decoder performance. Our work suggests how and why existing neural network decoders might be improved even further via data augmentation.

\section{Code availability}

All code and experiment data have been released in a public repository \cite{Peters_Code_for_Sample}. Numerics were done primarily using pytorch \cite{paszke2019pytorchimperativestylehighperformance}, stim \cite{Gidney_2021}, and pymatching \cite{higgott2023sparse}.

\section{Acknowledgments}
Thanks to Ewan Murphy, Zachary Mann, Vedran Dunjko, Moritz Lange, and Mats Granath for insightful discussions. This work was supported by the Applied Quantum Computing Challenge Program at the National Research Council of Canada. Research at Perimeter Institute is supported in part by the Government of Canada through the Department of Innovation, Science and Economic Development and by the Province of Ontario through the Ministry of Colleges and Universities. 

\clearpage

\printbibliography

@article{steane1996multiple,
  title={Multiple-particle interference and quantum error correction},
  author={Steane, Andrew},
  journal={Proceedings of the Royal Society of London. Series A: Mathematical, Physical and Engineering Sciences},
  volume={452},
  number={1954},
  pages={2551--2577},
  year={1996},
  publisher={The Royal Society London},
  doi={https://doi.org/10.1098/rspa.1996.0136}
}

@misc{senior2025scalablerealtimeneuraldecoder,
      title={A scalable and real-time neural decoder for topological quantum codes}, 
      author={Andrew W. Senior and Thomas Edlich and Francisco J. H. Heras and Lei M. Zhang and Oscar Higgott and James S. Spencer and Taylor Applebaum and Sam Blackwell and Justin Ledford and Akvilė Žemgulytė and Augustin Žídek and Noah Shutty and Andrew Cowie and Yin Li and George Holland and Peter Brooks and Charlie Beattie and Michael Newman and Alex Davies and Cody Jones and Sergio Boixo and Hartmut Neven and Pushmeet Kohli and Johannes Bausch},
      year={2025},
      eprint={2512.07737},
      archivePrefix={arXiv},
      primaryClass={quant-ph},
      url={https://arxiv.org/abs/2512.07737}, 
}

@article{Bravyi_2013,
   title={Simulation of rare events in quantum error correction},
   volume={88},
   ISSN={1094-1622},
   url={http://dx.doi.org/10.1103/PhysRevA.88.062308},
   DOI={10.1103/physreva.88.062308},
   number={6},
   journal={Physical Review A},
   publisher={American Physical Society (APS)},
   author={Bravyi, Sergey and Vargo, Alexander},
   year={2013},
   month=dec }

@misc{mayer2025rareeventsimulationquantum,
      title={Rare Event Simulation of Quantum Error-Correcting Circuits}, 
      author={Carolyn Mayer and Anand Ganti and Uzoma Onunkwo and Tzvetan Metodi and Benjamin Anker and Jacek Skryzalin},
      year={2025},
      eprint={2509.13678},
      archivePrefix={arXiv},
      primaryClass={quant-ph},
      url={https://arxiv.org/abs/2509.13678}, 
}

@MISC {2797618,
    TITLE = {(approximate) closed form for $\binom{n}{n/2}p^n$?},
    AUTHOR = {Claude Leibovici},
    HOWPUBLISHED = {Mathematics Stack Exchange},
    NOTE = {\url{https://math.stackexchange.com/q/2797618}},
}

@misc{Peters_Code_for_Sample,
author = {Peters, Evan},
title = {{Code for "Sample importance for data-driven decoding"}},
url = {https://github.com/peterse/mldec},
NOTE = {\url{https://github.com/peterse/mldec}},
}

@misc{paszke2019pytorchimperativestylehighperformance,
      title={PyTorch: An Imperative Style, High-Performance Deep Learning Library}, 
      author={Adam Paszke and Sam Gross and Francisco Massa and Adam Lerer and James Bradbury and Gregory Chanan and Trevor Killeen and Zeming Lin and Natalia Gimelshein and Luca Antiga and Alban Desmaison and Andreas Köpf and Edward Yang and Zach DeVito and Martin Raison and Alykhan Tejani and Sasank Chilamkurthy and Benoit Steiner and Lu Fang and Junjie Bai and Soumith Chintala},
      year={2019},
      eprint={1912.01703},
      archivePrefix={arXiv},
      primaryClass={cs.LG},
      url={https://arxiv.org/abs/1912.01703}, 
}

@article{valiant1984theory,
  title={A theory of the learnable},
  author={Valiant, Leslie G},
  journal={Communications of the ACM},
  volume={27},
  number={11},
  pages={1134--1142},
  year={1984},
  publisher={ACM New York, NY, USA}
}

@inproceedings{welling,
  author       = {Thomas N. Kipf and
                  Max Welling},
  title        = {Semi-Supervised Classification with Graph Convolutional Networks},
  booktitle    = {5th International Conference on Learning Representations, {ICLR} 2017,
                  Toulon, France, April 24-26, 2017, Conference Track Proceedings},
  year         = {2017},
  url          = {https://openreview.net/forum?id=SJU4ayYgl},
}

@article{google2023suppressing,
  title={Suppressing quantum errors by scaling a surface code logical qubit},
    author={Google Quantum AI},
  journal={Nature},
  volume={614},
  number={7949},
  pages={676--681},
  year={2023},
  publisher={Nature Publishing Group UK London}
}

@misc{wang_transformer_qec_2023,
	title = {Transformer-{QEC}: Quantum Error Correction Code Decoding with Transferable Transformers},
	url = {http://arxiv.org/abs/2311.16082},
	shorttitle = {Transformer-{QEC}},
	number = {{arXiv}:2311.16082},
	publisher = {{arXiv}},
	author = {Wang, Hanrui and Liu, Pengyu and Shao, Kevin and Li, Dantong and Gu, Jiaqi and Pan, David Z. and Ding, Yongshan and Han, Song},
	urldate = {2024-03-15},
	date = {2023-11-27},
	langid = {english},
	eprinttype = {arxiv},
	eprint = {2311.16082},
	% doi={https://doi.org/10.48550/arXiv.2311.16082},
}

@inproceedings{
choukroun_error_2022,
title={Error Correction Code Transformer},
author={Yoni Choukroun and Lior Wolf},
booktitle={Advances in Neural Information Processing Systems (Neurips)},
editor={Alice H. Oh and Alekh Agarwal and Danielle Belgrave and Kyunghyun Cho},
year={2022},
url={https://openreview.net/forum?id=4F0Pd2Wjl0}
}

@misc{cao_qecgpt_2023,
	title = {{qecGPT}: decoding Quantum Error-correcting Codes with Generative Pre-trained Transformers},
	url = {http://arxiv.org/abs/2307.09025},
	shorttitle = {{qecGPT}},
	number = {{arXiv}:2307.09025},
	publisher = {{arXiv}},
	author = {Cao, Hanyan and Pan, Feng and Wang, Yijia and Zhang, Pan},
	urldate = {2023-07-19},
	date = {2023-07-18},
	langid = {english},
	eprinttype = {arxiv},
	eprint = {2307.09025 [cond-mat, physics:quant-ph, stat]},
	keywords = {Quantum Physics, Computer Science - Machine Learning, Statistics - Machine Learning, Condensed Matter - Statistical Mechanics},
   % doi={https://doi.org/10.48550/arXiv.2307.09025},
}

@article{varsamopoulos_comparing_2020,
	title = {Comparing neural network based decoders for the surface code},
	volume = {69},
	issn = {0018-9340, 1557-9956, 2326-3814},
	url = {http://arxiv.org/abs/1811.12456},
	doi = {10.1109/TC.2019.2948612},
	pages = {300--311},
	number = {2},
	journaltitle = {{IEEE} Transactions on Computers},
	shortjournal = {{IEEE} Trans. Comput.},
	author = {Varsamopoulos, Savvas and Bertels, Koen and Almudever, Carmen G.},
	urldate = {2024-04-25},
	date = {2020-02-01},
	langid = {english},
	eprinttype = {arxiv},
	eprint = {1811.12456},
}

@article{chamberland_deep_2018,
	title = {Deep neural decoders for near term fault-tolerant experiments},
	volume = {3},
	issn = {2058-9565},
	url = {http://arxiv.org/abs/1802.06441},
	doi = {10.1088/2058-9565/aad1f7},
	pages = {044002},
	number = {4},
	journaltitle = {Quantum Science and Technology},
	shortjournal = {Quantum Sci. Technol.},
	author = {Chamberland, Christopher and Ronagh, Pooya},
	urldate = {2024-04-25},
	date = {2018-07-31},
	langid = {english},
	eprinttype = {arxiv},
	eprint = {1802.06441 [quant-ph, stat]},
}

@article{torlai_neural_2017,
	title = {Neural Decoder for Topological Codes},
	volume = {119},
	rights = {http://link.aps.org/licenses/aps-default-license},
	issn = {0031-9007, 1079-7114},
	url = {http://link.aps.org/doi/10.1103/PhysRevLett.119.030501},
	doi = {10.1103/PhysRevLett.119.030501},
	pages = {030501},
	number = {3},
	journaltitle = {Phys. Rev. Lett.},
	shortjournal = {Phys. Rev. Lett.},
	author = {Torlai, Giacomo and Melko, Roger G.},
	urldate = {2024-04-25},
	date = {2017-07-18},
	langid = {english},
}

@article{Choukroun_Wolf_2024, title={Deep Quantum Error Correction}, volume={38}, url={https://ojs.aaai.org/index.php/AAAI/article/view/27756}, DOI={10.1609/aaai.v38i1.27756}, abstractNote={Quantum error correction codes (QECC) are a key component for realizing the potential of quantum computing. QECC, as its classical counterpart (ECC), enables the reduction of error rates, by distributing quantum logical information across redundant physical qubits, such that errors can be detected and corrected. In this work, we efficiently train novel end-to-end deep quantum error decoders. We resolve the quantum measurement collapse by augmenting syndrome decoding to predict an initial estimate of the system noise, which is then refined iteratively through a deep neural network. The logical error rates calculated over finite fields are directly optimized via a differentiable objective, enabling efficient decoding under the constraints imposed by the code. Finally, our architecture is extended to support faulty syndrome measurement, by efficient decoding of repeated syndrome sampling. The proposed method demonstrates the power of neural decoders for QECC by achieving state-of-the-art accuracy, outperforming for small distance topological codes, the existing end-to-end neural and classical decoders, which are often computationally prohibitive.}, number={1}, journal={Proceedings of the AAAI Conference on Artificial Intelligence}, author={Choukroun, Yoni and Wolf, Lior}, year={2024}, month={3}, pages={64-72} }

@inproceedings{cammerer2022graph,
  title={Graph neural networks for channel decoding},
  author={Cammerer, Sebastian and Hoydis, Jakob and Aoudia, Fay{\c{c}}al A{\"\i}t and Keller, Alexander},
  booktitle={2022 IEEE Globecom Workshops (GC Wkshps)},
  pages={486--491},
  year={2022},
  organization={IEEE}
}

@inproceedings{
kwak2023boosting,
title={Boosting Learning for {LDPC} Codes to Improve the Error-Floor Performance},
author={Hee-Youl Kwak and Dae-Young Yun and Yongjune Kim and Sang-Hyo Kim and Jong-Seon No},
booktitle={Thirty-seventh Conference on Neural Information Processing Systems (Neurips)},
year={2023},
url={https://openreview.net/forum?id=Yj3lFEyfnl}
}

@inproceedings{KimJRKOV18,
  author={Hyeji Kim and Yihan Jiang and Ranvir Rana and Sreeram Kannan and Sewoong Oh and Pramod Viswanath},
  title={Communication Algorithms via Deep Learning},
  year={2018},
  cdate={1514764800000},
  url={https://openreview.net/forum?id=ryazCMbR-},
  booktitle={International Conference on Learning Representations (ICLR)},
}

@INPROCEEDINGS{8815400,
  author={Jiang, Yihan and Kannan, Sreeram and Kim, Hyeji and Oh, Sewoong and Asnani, Himanshu and Viswanath, Pramod},
  booktitle={2019 IEEE 20th International Workshop on Signal Processing Advances in Wireless Communications (SPAWC)}, 
  title={DEEPTURBO: Deep Turbo Decoder}, 
  year={2019},
  volume={},
  number={},
  pages={1-5},
  doi={10.1109/SPAWC.2019.8815400}}

@InProceedings{pmlr-v130-garcia-satorras21a,
  title = 	 { Neural Enhanced Belief Propagation on Factor Graphs },
  booktitle = {Proceedings of the International Conference on Artificial Intelligence and Statistics},
  author =       {Garcia Satorras, V{\'i}ctor and Welling, Max},
  pages = 	 {685--693},
  year = 	 {2021},
  volume = 	 {130},
  series = 	 {Proceedings of Machine Learning Research},
  month = 	 {4},
  publisher =    {PMLR},
  url = 	 {https://proceedings.mlr.press/v130/garcia-satorras21a.html}
}

@misc{maan2024machinelearningmessagepassingscalable,
      title={Machine Learning Message-Passing for the Scalable Decoding of QLDPC Codes}, 
      author={Arshpreet Singh Maan and Alexandru Paler},
      year={2024},
      eprint={2408.07038},
      archivePrefix={arXiv},
      primaryClass={quant-ph},
      url={https://arxiv.org/abs/2408.07038}, 
}

@article{nachmani2019hyper,
  title={Hyper-graph-network decoders for block codes},
  author={Nachmani, Eliya and Wolf, Lior},
  journal={Advances in Neural Information Processing Systems (Neurips)},
  volume={32},
  year={2019}
}

@inproceedings{ninkovic2024decodingquantumldpccodes,
  title={Decoding Quantum LDPC Codes Using Graph Neural Networks},
  author={Ninkovic, Vukan and Kundacina, Ognjen and Vukobratovic, Dejan and H{\"a}ger, Christian and i Amat, Alexandre Graell},
  booktitle={GLOBECOM 2024-2024 IEEE Global Communications Conference},
  pages={3479--3484},
  year={2024},
  organization={IEEE}
}

@misc{zhang_scalable_2023,
	title = {A Scalable, Fast and Programmable Neural Decoder for Fault-Tolerant Quantum Computation Using Surface Codes},
	url = {http://arxiv.org/abs/2305.15767},
	number = {{arXiv}:2305.15767},
	publisher = {{arXiv}},
	author = {Zhang, Mengyu and Ren, Xiangyu and Xi, Guanglei and Zhang, Zhenxing and Yu, Qiaonian and Liu, Fuming and Zhang, Hualiang and Zhang, Shengyu and Zheng, Yi-Cong},
	urldate = {2024-08-02},
	date = {2023-05-25},
	langid = {english},
	eprinttype = {arxiv},
	eprint = {2305.15767},
}

@article{davaasuren_general_2020,
	title = {General framework for constructing fast and near-optimal machine-learning-based decoder of the topological stabilizer codes},
	volume = {2},
	issn = {2643-1564},
	url = {https://link.aps.org/doi/10.1103/PhysRevResearch.2.033399},
	doi = {10.1103/PhysRevResearch.2.033399},
	pages = {033399},
	number = {3},
	journaltitle = {Phys. Rev. Research},
	shortjournal = {Phys. Rev. Research},
	author = {Davaasuren, Amarsanaa and Suzuki, Yasunari and Fujii, Keisuke and Koashi, Masato},
	urldate = {2024-08-02},
	date = {2020-09-11},
	langid = {english},
}

@article{lange_data_driven_2023,
  title = {Data-driven decoding of quantum error correcting codes using graph neural networks},
  author = {Lange, Moritz and Havstr\"om, Pontus and Srivastava, Basudha and Bengtsson, Isak and Bergentall, Valdemar and Hammar, Karl and Heuts, Olivia and van Nieuwenburg, Evert and Granath, Mats},
  journal = {Phys. Rev. Res.},
  volume = {7},
  issue = {2},
  pages = {023181},
  numpages = {14},
  year = {2025},
  month = {5},
  publisher = {American Physical Society},
  doi = {10.1103/PhysRevResearch.7.023181},
  url = {https://link.aps.org/doi/10.1103/PhysRevResearch.7.023181}
}

@article{ni_neural_2020,
   title={Neural Network Decoders for Large-Distance 2D Toric Codes},
   volume={4},
   ISSN={2521-327X},
   url={http://dx.doi.org/10.22331/q-2020-08-24-310},
   DOI={10.22331/q-2020-08-24-310},
   journal={Quantum},
   publisher={Verein zur Forderung des Open Access Publizierens in den Quantenwissenschaften},
   author={Ni, Xiaotong},
   year={2020},
   month=aug, pages={310} }

@article{overwater_neural_network_2022,
	title = {Neural-Network Decoders for Quantum Error Correction Using Surface Codes: A Space Exploration of the Hardware Cost-Performance Tradeoffs},
	volume = {3},
	issn = {2689-1808},
	url = {https://ieeexplore.ieee.org/document/9772289/?arnumber=9772289},
	doi = {10.1109/TQE.2022.3174017},
	shorttitle = {Neural-Network Decoders for Quantum Error Correction Using Surface Codes},
	pages = {1--19},
	journaltitle = {{IEEE} Transactions on Quantum Engineering},
	author = {Overwater, Ramon W. J. and Babaie, Masoud and Sebastiano, Fabio},
	urldate = {2024-08-02},
	date = {2022},
	note = {Conference Name: {IEEE} Transactions on Quantum Engineering},
}

@article{wagner_symmetries_2020,
	title = {Symmetries for a high-level neural decoder on the toric code},
	volume = {102},
	issn = {2469-9926, 2469-9934},
	url = {https://link.aps.org/doi/10.1103/PhysRevA.102.042411},
	doi = {10.1103/PhysRevA.102.042411},
	pages = {042411},
	number = {4},
	journaltitle = {Phys. Rev. A},
	shortjournal = {Phys. Rev. A},
	author = {Wagner, Thomas and Kampermann, Hermann and Bruß, Dagmar},
	urldate = {2024-08-02},
	date = {2020-10-26},
	langid = {english},
}

@article{meinerz_scalable_2022,
	title = {Scalable Neural Decoder for Topological Surface Codes},
	volume = {128},
	issn = {0031-9007, 1079-7114},
	url = {https://link.aps.org/doi/10.1103/PhysRevLett.128.080505},
	doi = {10.1103/PhysRevLett.128.080505},
	pages = {080505},
	number = {8},
	journaltitle = {Phys. Rev. Lett.},
	shortjournal = {Phys. Rev. Lett.},
	author = {Meinerz, Kai and Park, Chae-Yeun and Trebst, Simon},
	urldate = {2024-08-02},
	date = {2022-02-24},
	langid = {english},
}

@article{varsamopoulos_decoding_2018,
	title = {Decoding small surface codes with feedforward neural networks},
	volume = {3},
	issn = {2058-9565},
	url = {https://iopscience.iop.org/article/10.1088/2058-9565/aa955a},
	doi = {10.1088/2058-9565/aa955a},
	pages = {015004},
	number = {1},
	journaltitle = {Quantum Science and Technology},
	shortjournal = {Quantum Sci. Technol.},
	author = {Varsamopoulos, Savvas and Criger, Ben and Bertels, Koen},
	urldate = {2024-08-02},
	date = {2018-01},
	langid = {english},
}

@article{varbanov_neural_2023,
  title = {Neural network decoder for near-term surface-code experiments},
  author = {Varbanov, Boris M. and Serra-Peralta, Marc and Byfield, David and Terhal, Barbara M.},
  journal = {Phys. Rev. Res.},
  volume = {7},
  issue = {1},
  pages = {013029},
  numpages = {14},
  year = {2025},
  month = {1},
  publisher = {American Physical Society},
  doi = {10.1103/PhysRevResearch.7.013029},
  url = {https://link.aps.org/doi/10.1103/PhysRevResearch.7.013029}
}

@article{liu_neural_2019,
	title = {Neural Belief-Propagation Decoders for Quantum Error-Correcting Codes},
	volume = {122},
	issn = {0031-9007, 1079-7114},
	url = {https://link.aps.org/doi/10.1103/PhysRevLett.122.200501},
	doi = {10.1103/PhysRevLett.122.200501},
	pages = {200501},
	number = {20},
	journaltitle = {Phys. Rev. Lett.},
	shortjournal = {Phys. Rev. Lett.},
	author = {Liu, Ye-Hua and Poulin, David},
	urldate = {2024-04-25},
	date = {2019-05-22},
	langid = {english},
}

@article{5qubit2,
  title = {Mixed-state entanglement and quantum error correction},
  author = {Bennett, Charles H. and DiVincenzo, David P. and Smolin, John A. and Wootters, William K.},
  journal = {Phys. Rev. A},
  volume = {54},
  issue = {5},
  pages = {3824--3851},
  numpages = {0},
  year = {1996},
  month = {11},
  publisher = {American Physical Society},
  doi = {10.1103/PhysRevA.54.3824},
  url = {https://link.aps.org/doi/10.1103/PhysRevA.54.3824}
}

@article{5qubit1,
  title = {Perfect Quantum Error Correcting Code},
  author = {Laflamme, Raymond and Miquel, Cesar and Paz, Juan Pablo and Zurek, Wojciech Hubert},
  journal = {Phys. Rev. Lett.},
  volume = {77},
  issue = {1},
  pages = {198--201},
  numpages = {0},
  year = {1996},
  month = {7},
  publisher = {American Physical Society},
  doi = {10.1103/PhysRevLett.77.198},
  url = {https://link.aps.org/doi/10.1103/PhysRevLett.77.198}
}

@article{Massart_2006,
   title={Risk bounds for statistical learning},
   volume={34},
   ISSN={0090-5364},
   url={http://dx.doi.org/10.1214/009053606000000786},
   DOI={10.1214/009053606000000786},
   number={5},
   journal={The Annals of Statistics},
   publisher={Institute of Mathematical Statistics},
   author={Massart, Pascal and Nédélec, Élodie},
   year={2006},
   month=oct }

@article{bausch_learning_2023,
  title={Learning high-accuracy error decoding for quantum processors},
  author={Bausch, Johannes and Senior, Andrew W and Heras, Francisco JH and Edlich, Thomas and Davies, Alex and Newman, Michael and Jones, Cody and Satzinger, Kevin and Niu, Murphy Yuezhen and Blackwell, Sam and others},
  journal={Nature},
  pages={1--7},
  year={2024},
  publisher={Nature Publishing Group UK London},
  doi={https://doi.org/10.1038/s41586-024-08148-8},
}

@article{sweke_reinforcement_2021,
	title = {Reinforcement Learning Decoders for Fault-Tolerant Quantum Computation},
	volume = {2},
	issn = {2632-2153},
	url = {http://arxiv.org/abs/1810.07207},
	doi = {10.1088/2632-2153/abc609},
	pages = {025005},
	number = {2},
	journaltitle = {Machine Learning: Science and Technology},
	shortjournal = {Mach. Learn.: Sci. Technol.},
	author = {Sweke, Ryan and Kesselring, Markus S. and van Nieuwenburg, Evert P. L. and Eisert, Jens},
	urldate = {2024-04-25},
	date = {2021-06-01},
	langid = {english},
	eprinttype = {arxiv},
	eprint = {1810.07207},}

@article{PhysRevResearch.2.023230,
  title = {Deep Q-learning decoder for depolarizing noise on the toric code},
  author = {Fitzek, David and Eliasson, Mattias and Kockum, Anton Frisk and Granath, Mats},
  journal = {Phys. Rev. Res.},
  volume = {2},
  issue = {2},
  pages = {023230},
  numpages = {17},
  year = {2020},
  month = {5},
  publisher = {American Physical Society},
  doi = {10.1103/PhysRevResearch.2.023230},
  url = {https://link.aps.org/doi/10.1103/PhysRevResearch.2.023230}
}

@article{Andreasson_2019,
   title={Quantum error correction for the toric code using deep reinforcement learning},
   volume={3},
   ISSN={2521-327X},
   url={http://dx.doi.org/10.22331/q-2019-09-02-183},
   DOI={10.22331/q-2019-09-02-183},
   journal={Quantum},
   publisher={Verein zur Forderung des Open Access Publizierens in den Quantenwissenschaften},
   author={Andreasson, Philip and Johansson, Joel and Liljestrand, Simon and Granath, Mats},
   year={2019},
   month=9, pages={183} }

@article{krastanov_deep_2017,
	title = {Deep Neural Network Probabilistic Decoder for Stabilizer Codes},
	volume = {7},
	issn = {2045-2322},
	url = {http://arxiv.org/abs/1705.09334},
	doi = {10.1038/s41598-017-11266-1},
	pages = {11003},
	number = {1},
	journaltitle = {Scientific Reports},
	shortjournal = {Sci Rep},
	author = {Krastanov, Stefan and Jiang, Liang},
	urldate = {2024-05-07},
	date = {2017-09-08},
	langid = {english},
	eprinttype = {arxiv},
	eprint = {1705.09334},
}

@article{breuckmann_scalable_2018,
	title = {Scalable Neural Network Decoders for Higher Dimensional Quantum Codes},
	volume = {2},
	issn = {2521-327X},
	url = {http://arxiv.org/abs/1710.09489},
	doi = {10.22331/q-2018-05-24-68},
	pages = {68},
	journaltitle = {Quantum},
	shortjournal = {Quantum},
	author = {Breuckmann, Nikolas P. and Ni, Xiaotong},
	urldate = {2024-05-07},
	date = {2018-05-24},
	langid = {english},
	eprinttype = {arxiv},
	eprint = {1710.09489},
}

@article{gicev_scalable_2023,
	title = {A scalable and fast artificial neural network syndrome decoder for surface codes},
	volume = {7},
	issn = {2521-327X},
	url = {http://arxiv.org/abs/2110.05854},
	doi = {10.22331/q-2023-07-12-1058},
	pages = {1058},
	journaltitle = {Quantum},
	shortjournal = {Quantum},
	author = {Gicev, Spiro and Hollenberg, Lloyd C. L. and Usman, Muhammad},
	urldate = {2024-03-15},
	date = {2023-07-12},
	langid = {english},
	eprinttype = {arxiv},
	eprint = {2110.05854},
}

@article{maskara_advantages_2019,
	title = {Advantages of versatile neural-network decoding for topological codes},
	volume = {99},
	issn = {2469-9926, 2469-9934},
	url = {https://link.aps.org/doi/10.1103/PhysRevA.99.052351},
	doi = {10.1103/PhysRevA.99.052351},
	pages = {052351},
	number = {5},
	journaltitle = {Phys. Rev. A},
	shortjournal = {Phys. Rev. A},
	author = {Maskara, Nishad and Kubica, Aleksander and Jochym-O'Connor, Tomas},
	urldate = {2024-05-06},
	date = {2019-05-30},
	langid = {english},
}

@misc{matekole_decoding_2022,
	title = {Decoding surface codes with deep reinforcement learning and probabilistic policy reuse},
	url = {http://arxiv.org/abs/2212.11890},
	number = {{arXiv}:2212.11890},
	publisher = {{arXiv}},
	author = {Matekole, Elisha Siddiqui and Ye, Esther and Iyer, Ramya and Chen, Samuel Yen-Chi},
	urldate = {2024-04-25},
	date = {2022-12-22},
	langid = {english},
	eprinttype = {arxiv},
	eprint = {2212.11890},
}

@article{PhysRevResearch.6.L032004,
  title = {Artificial neural network syndrome decoding on IBM quantum processors},
  author = {Hall, Brhyeton and Gicev, Spiro and Usman, Muhammad},
  journal = {Phys. Rev. Res.},
  volume = {6},
  issue = {3},
  pages = {L032004},
  numpages = {7},
  year = {2024},
  month = {7},
  publisher = {American Physical Society},
  doi = {10.1103/PhysRevResearch.6.L032004},
  url = {https://link.aps.org/doi/10.1103/PhysRevResearch.6.L032004}
}

@article{sivak2024optimizationdecoderpriorsaccurate,
  title={Optimization of decoder priors for accurate quantum error correction},
  author={Sivak, Volodymyr and Newman, Michael and Klimov, Paul},
  journal={Phys. Rev. Lett.},
  volume={133},
  number={15},
  pages={150603},
  year={2024},
  publisher={APS}
}

@article{heLearningImbalancedData2009,
	title = {Learning from Imbalanced Data},
	volume = {21},
	issn = {1558-2191},
	url = {https://ieeexplore.ieee.org/document/5128907/?arnumber=5128907},
	doi = {10.1109/TKDE.2008.239},
	pages = {1263--1284},
	number = {9},
	journaltitle = {{IEEE} Transactions on Knowledge and Data Engineering},
	author = {He, Haibo and Garcia, Edwardo A.},
	urldate = {2025-02-20},
	date = {2009-09},
}

@article{japkowicz2002class,
  title={The class imbalance problem: A systematic study},
  author={Japkowicz, Nathalie and Stephen, Shaju},
  journal={Intelligent data analysis},
  volume={6},
  number={5},
  pages={429--449},
  year={2002},
  publisher={IOS Press}
}

@misc{freedman1998projectiveplaneplanarquantum,
      title={Projective plane and planar quantum codes}, 
      author={Michael H. Freedman and David A. Meyer},
      year={1998},
      eprint={quant-ph/9810055},
      archivePrefix={arXiv},
      primaryClass={quant-ph},
      url={https://arxiv.org/abs/quant-ph/9810055}, 
}

@misc{bravyi1998quantumcodeslatticeboundary,
      title={Quantum codes on a lattice with boundary}, 
      author={S. B. Bravyi and A. Yu. Kitaev},
      year={1998},
      eprint={quant-ph/9811052},
      archivePrefix={arXiv},
      primaryClass={quant-ph},
      url={https://arxiv.org/abs/quant-ph/9811052}, 
}

@misc{higgott2023sparse,
  title={Sparse Blossom: correcting a million errors per core second with minimum-weight matching},
  author={Higgott, Oscar and Gidney, Craig},
  eprint={2303.15933},
  archivePrefix={arXiv},
  year={2023}
}

@article{Gidney_2021,
   title={Stim: a fast stabilizer circuit simulator},
   volume={5},
   ISSN={2521-327X},
   url={http://dx.doi.org/10.22331/q-2021-07-06-497},
   DOI={10.22331/q-2021-07-06-497},
   journal={Quantum},
   publisher={Verein zur Forderung des Open Access Publizierens in den Quantenwissenschaften},
   author={Gidney, Craig},
   year={2021},
   month=7, pages={497} }

@article{krawczyk2016learning,
  title={Learning from imbalanced data: open challenges and future directions},
  author={Krawczyk, Bartosz},
  journal={Progress in artificial intelligence},
  volume={5},
  number={4},
  pages={221--232},
  year={2016},
  publisher={Springer}
}

@inproceedings{bengio2009curriculum,
  title={Curriculum learning},
  author={Bengio, Yoshua and Louradour, J{\'e}r{\^o}me and Collobert, Ronan and Weston, Jason},
  booktitle={Proceedings of the 26th annual international conference on machine learning (ICML)},
  pages={41--48},
  year={2009}
}

@ARTICLE{wang2022,
  author={Wang, Xin and Chen, Yudong and Zhu, Wenwu},
  journal={IEEE Transactions on Pattern Analysis and Machine Intelligence}, 
  title={A Survey on Curriculum Learning}, 
  year={2022},
  volume={44},
  number={9},
  pages={4555-4576},
  doi={10.1109/TPAMI.2021.3069908}}

@article{sovianyCurriculumLearningSurvey2022,
  title={Curriculum learning: A survey},
  author={Soviany, Petru and Ionescu, Radu Tudor and Rota, Paolo and Sebe, Nicu},
  journal={International Journal of Computer Vision},
  volume={130},
  number={6},
  pages={1526--1565},
  year={2022},
  publisher={Springer}
}

@article{angluin1988learning,
  title={Learning from noisy examples},
  author={Angluin, Dana and Laird, Philip},
  journal={Machine learning},
  volume={2},
  pages={343--370},
  year={1988},
  publisher={Springer}
}

@article{arratia1989tutorial,
  title={Tutorial on large deviations for the binomial distribution},
  author={Arratia, Richard and Gordon, Louis},
  journal={Bulletin of mathematical biology},
  volume={51},
  number={1},
  pages={125--131},
  year={1989},
  publisher={Springer}
}

@article{vaswani2017attention,
  title={Attention is all you need},
  author={Vaswani, Ashish and Shazeer, Noam and Parmar, Niki and Uszkoreit, Jakob and Jones, Llion and Gomez, Aidan N and Kaiser, {\L}ukasz and Polosukhin, Illia},
  journal={Advances in neural information processing systems (Neurips)},
  volume={30},
  year={2017}
}

@article{li2017efficient,
  title={Efficient variational quantum simulator incorporating active error minimization},
  author={Li, Ying and Benjamin, Simon C},
  journal={Phys. Rev. X},
  volume={7},
  number={2},
  pages={021050},
  year={2017},
  publisher={APS}
}

@article{temme2017error,
  title={Error mitigation for short-depth quantum circuits},
  author={Temme, Kristan and Bravyi, Sergey and Gambetta, Jay M},
  journal={Phys. Rev. Lett.},
  volume={119},
  number={18},
  pages={180509},
  year={2017},
  publisher={APS}
}

@article{kandala2019error,
  title={Error mitigation extends the computational reach of a noisy quantum processor},
  author={Kandala, Abhinav and Temme, Kristan and C{\'o}rcoles, Antonio D and Mezzacapo, Antonio and Chow, Jerry M and Gambetta, Jay M},
  journal={Nature},
  volume={567},
  number={7749},
  pages={491--495},
  year={2019},
  publisher={Nature Publishing Group UK London}
}

@article{varsamopoulos2019decodingsurfacecodedistributed,
  title={Decoding surface code with a distributed neural network--based decoder},
  author={Varsamopoulos, Savvas and Bertels, Koen and Almudever, Carmen G},
  journal={Quantum Machine Intelligence},
  volume={2},
  pages={1--12},
  year={2020},
  publisher={Springer},
  doi={https://doi.org/10.1007/s42484-020-00015-9}
}

@article{baireuther_machine_2018,
	title = {Machine-learning-assisted correction of correlated qubit errors in a topological code},
	volume = {2},
	issn = {2521-327X},
	url = {http://arxiv.org/abs/1705.07855},
	doi = {10.22331/q-2018-01-29-48},
	pages = {48},
	journaltitle = {Quantum},
	shortjournal = {Quantum},
	author = {Baireuther, P. and O'Brien, T. E. and Tarasinski, B. and Beenakker, C. W. J.},
	urldate = {2024-04-26},
	date = {2018-01-29},
	langid = {english},
	eprinttype = {arxiv},
	eprint = {1705.07855 [cond-mat, physics:quant-ph]},
}

@misc{bhoumikEfficientDecodingSurface2021,
	title = {Efficient Decoding of Surface Code Syndromes for Error Correction in Quantum Computing},
	url = {http://arxiv.org/abs/2110.10896},
	number = {{arXiv}:2110.10896},
	publisher = {{arXiv}},
	author = {Bhoumik, Debasmita and Sen, Pinaki and Majumdar, Ritajit and Sur-Kolay, Susmita and J, Latesh Kumar K. and Iyengar, Sundaraja Sitharama},
	urldate = {2024-10-17},
	date = {2021-10-21},
	langid = {english},
	eprinttype = {arxiv},
	eprint = {2110.10896},
    % doi={https://doi.org/10.48550/arXiv.2110.10896}
}

@misc{iyerHardnessDecodingQuantum2013,
	title = {Hardness of decoding quantum stabilizer codes},
	url = {http://arxiv.org/abs/1310.3235},
	number = {{arXiv}:1310.3235},
	publisher = {{arXiv}},
	author = {Iyer, Pavithran and Poulin, David},
	urldate = {2024-10-02},
	date = {2013-10-11},
	langid = {english},
	eprinttype = {arxiv},
	eprint = {1310.3235},
    doi={https://doi.org/10.48550/arXiv.1310.3235},
}

@INPROCEEDINGS{kuo_lu_2012,
  author={Kuo, Kao-Yueh and Lu, Chung-Chin},
  booktitle={2012 International Symposium on Information Theory and its Applications}, 
  title={On the hardness of decoding quantum stabilizer codes under the depolarizing channel}, 
  year={2012},
  volume={},
  number={},
  pages={208-211},
  doi={}}

@article{hsiehNPhardnessDecodingQuantum2011,
	title = {{NP}-hardness of decoding quantum error-correction codes},
	volume = {83},
	rights = {http://link.aps.org/licenses/aps-default-license},
	issn = {1050-2947, 1094-1622},
	url = {https://link.aps.org/doi/10.1103/PhysRevA.83.052331},
	doi = {10.1103/PhysRevA.83.052331},
	pages = {052331},
	number = {5},
	journaltitle = {Phys. Rev. A},
	shortjournal = {Phys. Rev. A},
	author = {Hsieh, Min-Hsiu and Le Gall, François},
	urldate = {2024-10-04},
	date = {2011-05-31},
	langid = {english},
}

@article{chamberlandTechniquesCombiningFast2023,
	title = {Techniques for combining fast local decoders with global decoders under circuit-level noise},
	volume = {8},
	issn = {2058-9565},
	url = {http://arxiv.org/abs/2208.01178},
	doi = {10.1088/2058-9565/ace64d},
	pages = {045011},
	number = {4},
	journaltitle = {Quantum Science and Technology},
	shortjournal = {Quantum Sci. Technol.},
	author = {Chamberland, Christopher and Goncalves, Luis and Sivarajah, Prasahnt and Peterson, Eric and Grimberg, Sebastian},
	urldate = {2024-10-17},
	date = {2023-10-01},
	langid = {english},
	eprinttype = {arxiv},
	eprint = {2208.01178},
}

@misc{egorovENDEquivariantNeural2023,
	title = {The {END}: An Equivariant Neural Decoder for Quantum Error Correction},
	url = {http://arxiv.org/abs/2304.07362},
	shorttitle = {The {END}},
	number = {{arXiv}:2304.07362},
	publisher = {{arXiv}},
	author = {Egorov, Evgenii and Bondesan, Roberto and Welling, Max},
	urldate = {2024-10-30},
	date = {2023-04-14},
	langid = {english},
	eprinttype = {arxiv},
	eprint = {2304.07362},
    % doi={https://doi.org/10.48550/arXiv.2304.07362},
}

@inproceedings{gcn_ref,
author = {Morris, Christopher and Ritzert, Martin and Fey, Matthias and Hamilton, William L. and Lenssen, Jan Eric and Rattan, Gaurav and Grohe, Martin},
title = {Weisfeiler and leman go neural: higher-order graph neural networks},
year = {2019},
isbn = {978-1-57735-809-1},
doi = {10.1609/aaai.v33i01.33014602},
booktitle = {Proceedings of the Thirty-Third AAAI Conference on Artificial Intelligence},
articleno = {565},
numpages = {8},
location = {Honolulu, Hawaii, USA},
series = {AAAI'19}
}

@article{frenay2013classification,
  title={Classification in the presence of label noise: a survey},
  author={Frenay, Benoit and Verleysen, Michel},
  journal={IEEE transactions on neural networks and learning systems},
  volume={25},
  number={5},
  pages={845--869},
  year={2013},
  publisher={IEEE}
}

@inproceedings{xiao2015learning,
  title={Learning from massive noisy labeled data for image classification},
  author={Xiao, Tong and Xia, Tian and Yang, Yi and Huang, Chang and Wang, Xiaogang},
  booktitle={Proceedings of the IEEE conference on computer vision and pattern recognition},
  pages={2691--2699},
  year={2015}
}

@misc{song2022learningnoisylabelsdeep,
      title={Learning from Noisy Labels with Deep Neural Networks: A Survey}, 
      author={Hwanjun Song and Minseok Kim and Dongmin Park and Yooju Shin and Jae-Gil Lee},
      year={2022},
      eprint={2007.08199},
      archivePrefix={arXiv},
      primaryClass={cs.LG},
      url={https://arxiv.org/abs/2007.08199}, 
}

\newpage
\appendix

\section{Machine learning for decoding simple (Q)ECCs}\label{app:background}
Whenever possible, we denote by (capital letter) $A$ a random variable taking on values (lowercase) $a$ from a finite set (mathcal) $\mathcal{A}$ with probability $p_A(a)$, where $p_A: \mathcal{A} \rightarrow \mathbb{R}^+$ denotes a discrete probability distribution. We write the indicator function as $\I\{ \cdot\}$, e.g. $\I\{a =1\}$ is one if $a=1$ and zero otherwise. $\Pr_A(\cdot)$ or $\Pr_{a \sim p_A}(\cdot)$ denote probabilities of some event with respect to $p_A$, e.g. $\Pr_A(A=1) = \sum_{a \in \mathcal{A}}p_A(a)\I\{a=1\}$. The expectation value $\E$ is a linear operator on functions of random variables, with $\E[f(A)] := \sum_{a\in\mathcal{A}} p_A(a) f(a)$; a useful identity is $\Pr_A(\cdot) = \E_A[\I\{\cdot\}]$.

For completeness, we summarize the stabilizer formalism for QECCs. Define the Pauli group on $n$ qubits as the group generated by $n$-qubit Paulis mod phase,  $\mPhat_n := \langle\{I, X, Y, Z\}^{\otimes n} \rangle / \{\pm I, \pm iI\}$. We are given a stabilizer code with generators 
$$G_\mS:=\{g_1, \dots, g_r\}\subset \mP_n$$

generating the abelian subgroup $S = \langle g_1, \dots, g_r\rangle$ such that $-I \notin \mS$. The code is equipped with a logical Pauli group $\mL := \Nhat(\mS)/\mS$, where $\Nhat(\mS)$ denotes the centralizer of $S$ (mod $\{\pm I, \pm iI\}$). Furthermore, we can specify a set of canonical errors $ G_\mT:= \{h_1, \dots, h_r\}$ such that $\{h_i, g_i\}=0$ and $[h_i, g_j] = 0$ for $j\neq i$, which generates the group of \textit{pure errors}  $T:=\langle h_1, \dots, h_r\rangle$. For an $[[n, k, d]]$ code, where $r = n-k$, we then have a partitioning 
\begin{align}\label{eq:prism}
    \mPhat_n = \mL \times \mS   \times \mT
\end{align}
where $|\mL| = 2^{2k}$, $|\mS| = |\mT| = 2^{n-k}$, in the sense that each Pauli operator decomposes as $f = s\ell t$ for $( \ell, s, t) \in \mL \times \mS \times \mT $. A choice of probability distribution for errors $p_E$ defined over $\mPhat$ induces a joint distribution $p_{LST}$ over this partition defined by $p_{LST}(\ell(e), s(e), t(e)) := p_E(e)$. Conditional distributions are denoted like $p_{L|T}$, and we will abuse this notation when the meaning is obvious, such as $p_{ET}(e,t):= p_E(e) \I\{t = t(e)\}$.

Before describing decoding, we briefly review degeneracy in QECCs. For a set of errors $\mE \subseteq \mPhat$, we say that a code is called \textit{degenerate with respect to} $\mE$ if two errors differ by a stabilizer only, i.e. $e^\dagger f \in \mS$ for some $e,f \in \mE$. A code is \textit{nondegenerate with respect to} $\mE$ if it is not degenerate with respect to $\mE$. Usually it is assumed that $\mE$ is a set of correctable errors for the code, e.g. for a code with a distance $d = 2t+1$ this means $\mE = \mE^{(t)}:= \{e \in \mPhat: \wt(e) \leq t\}$. In this case, a nondegenerate code means that no two correctable errors share a syndrome, i.e. $ e,f\in\mE^{(t)} \Rightarrow e^\dagger f \notin N(\mS)$.\footnote{There is some ambiguity about the definition of degeneracy with respect to a set of errors that is not correctable: Either non-degenerate means all $e,f \in \mE$  have different syndromes or degenerate means $e \in f\mS$ for some $e,f \in \mE$. These are not equivalent in general.} However, we will often consider decoding general Pauli errors i.e. $\mE = \mPhat_n$, in which case every code is degenerate. Now, given a Pauli error $e = s(e) \ell(e) t(e) \in \mPhat$, its \textit{syndrome} $\sigma(e) \in \{0,1\}^{n-k}$ is a bitstring that identifies the corresponding pure error $t(e)$, such that the $i$-th bit in the syndrome $\sigma(e)$ is $1$ if $\{E, h_i\}=1$ and is otherwise $0$. We will denote the pure error corresponding to $\sigma$ as $t_\sigma$. The syndrome can be measured via appropriate stabilizer measurements, and its joint distribution with logical operators is $p_{L\Sigma}$ while its joint distribution with errors is $p_{\Sigma E}$. We can now consider two distinct decoding problems:

\begin{itemize}
    \item \textit{Degenerate decoding}: Given a syndrome $\sigma(e)$, output a label for the logical operator $\ell(e)$
    \item \textit{Nondegenerate decoding}: Given a syndrome $\sigma(e)$, output the error $e$.
\end{itemize}

In both cases, the stabilizer component $s(e)$ does not matter for implementing corrections. In the nondegenerate problem, the mappings $\ell: \mE \rightarrow \mL$ and $\sigma: \mE \rightarrow \{0,1\}^{n-k}$ are invertible on the set of correctable errors, so that determining either of $\ell(e)$ and $e$ is equivalent. In the degenerate problem, $\sigma(e)$ does not uniquely correspond to $\ell(e)$ and multiple (correctable) errors may share a syndrome. A decoder that fails at nondegenerate decoding may succeed at degenerate decoding: Given $\sigma(e)$, such a decoder may output $e' \neq e$ that satisfies $e' \in \ell(e)\mS$. For brevity, we may say that a decoder ``successfully decodes an error $e$'' when the decoder correctly decodes $\sigma(e)$ in either the degenerate or nondegenerate sense. The logical error probability (LEP) is the total probability that a decoder fails with respect to errors sampled from $p_E$.

Given a syndrome as input, a nondegenerate maximum likelihood decoder $\MLD$ is a decoder that outputs the most likely error to have occurred, and a degenerate $\MLD$ outputs the most likely logical error to have occurred. We denote the $\MLD$s by $f^*$, defined according to
\begin{align}
    f^*(\sigma) &\in \argmax_{\ell \in \mL} p_{L|\Sigma} (\ell|\sigma) \qquad \text{(degenerate)}\label{eq:dqmld} \\
    f^*(\sigma) &\in \argmax_{e \in \mPhat} p_{E|\Sigma} (e|\sigma) \qquad \text{(nondegenerate)} \label{eq:qmld}
\end{align}
such that, in the event of ties \textit{any} solution maximizing Eq.~\ref{eq:dqmld} or Eq.~\ref{eq:qmld} is considered maximum likelihood. The distinction between degenerate and nondegenerate $\MLD$ will be clear from context, and we will sometimes use $\MLD(\sigma)$ to denote the \textit{entire set} of maximum likelihood outputs $\{f^*(\sigma)\}$ satisfying Eqs.~\ref{eq:dqmld}-\ref{eq:qmld}

\section{Example importance}\label{app:importance}

We provide additional details on example importance. First, we show that the cumulative importance of a sample provides an upper bound for the difference in generalization accuracy between any learning model $\hat{f}$ and a fixed baseline $f_0$, as reported in Eq.~\ref{eq:imp_ub}:
\begin{align}
    \Pr_{(\sigma,y)\sim \mD} &(\hat{f}(\sigma) = y) - \Pr_{(\sigma,y)\sim \mD} (f_0(\sigma) = y) 
    \\&\leq \Pr_{(\sigma,y)\sim \mD} (f^*(\sigma) = y) - \Pr_{(\sigma,y)\sim \mD} (f_0(\sigma) = y)\label{line:a1}
    \\&= \Pr_{(\sigma,y)\sim \mD} (f^*(\sigma) = y)  
    \\\nonumber &\qquad-\left(\Pr_{(\sigma,y)\sim \mD} (f_0(\sigma) = y, f^*(\sigma) =y) + \Pr_{(\sigma,y)\sim \mD} (f_0(\sigma) = y, f^*(\sigma) \neq y) \right)
    \\&= \Pr_{(\sigma,y)\sim \mD} (f^*(\sigma) = y, f_0(\sigma) \neq y) - \Pr_{(\sigma,y)\sim \mD} (f_0(\sigma) = y, f^*(\sigma) \neq y) 
    \\&= \E_{(\sigma,y)\sim \mD} [\I\{f^*(\sigma)=y\}\I\{f_0(\sigma)\neq y\}]  - \underbrace{\Pr_{(\sigma,y)\sim \mD} (f_0(\sigma) = y, f^*(\sigma) \neq y) }_{\textsc{gap}}
    \\&\leq \sum_{(\sigma,y)\sim \mD} J((\sigma,y); f_0)   
\end{align}
The first bound is the gap between an NN decoder and optimal decoding, which can be saturated but we don't expect it to be, and the final line is the cumulative importance of the data. The difference between the last two lines, \textsc{gap}, is the total probability of errors where $\MLD$ is wrong but the baseline is correct (unlikely).

Recall that for some syndromes, there may be several possible choices of $f^*$ that satisfy the $\MLD$ rule, e.g. when $(\sigma, y)$ and $(\sigma, y')$ where $y, y'$ are both $\MLD$ by Eqs.~\ref{eq:dqmld}-\ref{eq:qmld}. We have defined importance such that only one such example is important, while the other is  ``bad''. Thus, when two examples each pair $\sigma$ with some $\MLD$ decoding rule, this represents another form of ``label noise''. Note that different choices of $f^*$ may affect the total importance of a sample $D_N$, but will not affect the total importance of a distribution $p_E$. Choosing different $f^*$ only has the effect of assigning a different (equal probability) subset to be important. 

Another consequence of this is that the importance of a sample might be much smaller than the cumulative probability of examples $(\sigma, y)$ where $y$ is a $\MLD$ label for $\sigma$; in this sense, the plot in Fig.~\ref{fig:toric_results}a is undercounting important examples. However, since decoding accuracy cannot be improved by switching between different $\MLD$ decoders, Eq.~\ref{eq:imp_ub} still holds and ultimately this justifies the definition of example importance.

\section{Analysis of repetition code}\label{app:toymodel}

In this appendix we provide more detail on statements from the main text, as well as proofs for Proposition~\ref{prop:1} and Corollary~\ref{cor:1}.

A common theme in example importance is that higher weight errors are less likely. In the main text, we have frequently used Bernstein's inequality to make this intuition rigorous. For instance, in an error model with error distribution $p_E$ describing $n$ bits each experiencing independent bitflips with probability at most $p$, we have
\begin{align}\label{eq:bernstein}
    \Pr_{p_{E}}(\wt(E) \geq t) \leq \exp \left(-\frac{t^2}{np + t/3} \right) \leq \exp(-t^2/np)
\end{align}
In particular, for $t=n/2$, we have $\Pr(\wt(E) \geq n/2) \leq \exp(-n/4p)$.

\subsection{Proof of Proposition~\ref{prop:1}: \texorpdfstring{$\DDD$}{DDD} for the classical repetition code is noisy binary classification}\label{app:prop_1}

We show that $\DDD$ with a $k=1$ classical repetition code is equivalent to noisy binary classification, for any fixed, known choice of $f_0$. One way to demonstrate this equivalence is to show that the dataset provided $\DDD$ is equivalent to some other dataset with noisy, binary labels. Thus, our goal is to build an oracle for noisy sampling that can be sampled to produce datasets $D_N$ that are indistinguishable from those provided in the $\DDD$ problem.

An instance of $\DDD$ for the repetition code involves a dataset $D_N = \{(\sigma_i, e_i)\}_{i=1}^N$ where each syndrome is computed $\sigma_i = H e_i$ for the parity check matrix $H$ of the classical code. Writing $1^n := (1, \dots, 1)^T$ as the all-ones vector, we see that $H1^n=0$ and therefore every syndrome $\sigma_i = He_i$ of an error $e_i$ also satisfies $\sigma_i = H(e_i + 1^n)$. 

We can therefore relate $\DDD$ to the binary classification task of answering the question ``given $\sigma_i$, should I return $e_i$ or $e_i \oplus 1^n$?''. However, in order to pose this question we first need to generate a guess for $e_i$, and then decide whether to reject this guess because it is the less likely of the two possibilities. The role of $f_0$ in Proposition~\ref{prop:1} is therefore to provide an initial guess $e_i = f_0(\sigma_i)$, and the classification task is, given $\sigma$, to output a label deciding whether to accept ($z=1$) or reject ($z=0$):
\begin{equation}\label{eq:noisy_label}
    z(\sigma) := \I\{f_0(\sigma) = f^*(\sigma)\}.
\end{equation}
For a fixed $\MLD$ $f^*$, we can partition the set of possible examples in half:
\begin{equation}
    \{(\sigma, e): e \in \{0,1\}^n, \sigma=He\} = \{(\sigma, e): e= f^*(\sigma)\} \cup\{(\sigma, e): e\neq f^*(\sigma)\}
\end{equation}
The first set is all good examples, the second set is all bad examples. For a given syndrome $\sigma$, define the probability that the corresponding $(\sigma, e)$ is a bad example as
\begin{align}
    \eta(\sigma) :&= \Pr_{e \sim p_E}(He=\sigma, (\sigma, e)\text{ is bad} )
        \\&= \Pr_{e \sim p_E}(He=\sigma,e \neq f^*(He)).
\end{align}
In other words, an example $(\sigma, e)$ sampled from  $p_{\Sigma E}$ obeys
\begin{align}\label{eq:ddd_noise0}
    \Pr_{e}((\sigma, e) = (\sigma, f^*(\sigma))) &= p_\Sigma(\sigma) - \eta(\sigma) \\
    \Pr_{e}((\sigma, e) \neq (\sigma, f^*(\sigma))) &=  \eta(\sigma), \label{eq:ddd_noise}
\end{align}
where $p_\Sigma$ is the distribution for syndromes induced by $p_E$ and $H$. Now we construct a noisy examples oracle $EX^{\vec{\eta}}$ which outputs $\sigma$ alongside a binary label and show how this oracle can be sampled to build $D_N$. The oracle generates an example by the following procedure:
\begin{enumerate}
    \item Sample $ \sigma \sim p_{\Sigma}$
    \item Sample from a noisy $\MLD$ source $\tilde{g}$  that outputs
    \begin{equation}
        \tilde{g}(\sigma) = \begin{cases}
            f^*(\sigma)& \Pr=1 - \eta(\sigma)/p_\Sigma(\sigma) \\
            f^*(\sigma) \oplus 1^n, & \Pr = \eta(\sigma)/p_\Sigma(\sigma)
        \end{cases}
    \end{equation}
    \item Label $\tilde{z}(\sigma) =\I\{f_0(\sigma) = \tilde{g}(\sigma)\}$
\end{enumerate}
Observe that the distribution $(\sigma, \tilde{g}(\sigma))$ at step 2 of this procedure is identical to the sampling scheme described by Eqs.~\ref{eq:ddd_noise0}-\ref{eq:ddd_noise}, and step 3 just requires the use of $f_0$. This means that generating datasets according to $\DDD$ and using $f_0$ for postprocessing can be used to simulate $EX^{\vec{\eta}}$; i.e. sampling from $\DDD$ given $f_0$ reduces to sampling from $EX^{\vec{\eta}}$.

We can recognize the label $\tilde{z}(\sigma)$ as a \textit{noisy} version of our desired label $z(\sigma)$ in Eq.~\ref{eq:noisy_label}. The noise rate is \textit{instance dependent}, as each noise probability depends on $\sigma$. However, we can compute the total probability of noise over all syndromes as
\begin{align}
    \Pr_{\sigma,\tilde{z}}(\tilde{z}(\sigma) \neq z(\sigma)) &=  \Pr_{\sigma,\tilde{z}} (\tilde{z}(\sigma) = 1 | z (\sigma) = 0) + \Pr_{\sigma,\tilde{z}} (\tilde{z}(\sigma) = 0 | z (\sigma) = 1)
    \\ &= \Pr_{\tilde{g},\sigma:f_0(\sigma)\neq f^*(\sigma)}(f_0(\sigma) = \tilde{g}(\sigma) ) + \Pr_{\tilde{g},\sigma:f_0(\sigma)= f^*(\sigma)}(f_0(\sigma) \neq \tilde{g}(\sigma) )
    \\&= \Pr_{\tilde{g},\sigma:f_0(\sigma)\neq f^*(\sigma)}(f^*(\sigma) \neq \tilde{g}(\sigma) ) + \Pr_{\tilde{g},\sigma:f_0(\sigma)= f^*(\sigma)}(f^*(\sigma) \neq \tilde{g}(\sigma) )
    \\&= \sum_\sigma \eta(\sigma)
    \\&= \Pr(\text{bad)}
\end{align}
Meanwhile, the total probability of each class can be computed as
\begin{align}
    \Pr_\sigma(z(\sigma)=0) &= \Pr_\sigma(f^*(\sigma) \neq f_0(\sigma))
    \\&= \Pr_{\sigma,e}(f^*(\sigma) \neq f_0(\sigma), f^*(\sigma) = e) + \Pr_{\sigma,e}(f^*(\sigma) \neq f_0(\sigma), f^*(\sigma) \neq e)
    \\&\leq \Pr(\text{important}) + \Pr(\text{bad})
\\
    \Pr_\sigma(z(\sigma)=1) &= \Pr_\sigma(f^*(\sigma) = f_0(\sigma))
    \\&= \Pr_{\sigma,e}(f^*(\sigma) = f_0(\sigma), f^*(\sigma) = e) + \Pr_{\sigma,e}(f^*(\sigma) = f_0(\sigma), f^*(\sigma) \neq e)
    \\&\leq \Pr(\text{unimportant}) + \Pr(\text{bad})
\end{align}
In both cases, we can bound the probability of bad examples, since errors with weight less than $\frac{d}{2}$ must be correctable (and therefore good). A standard Hoeffding bound for an iid error model with individual error rates $p$ gives
\begin{align}\label{eq:dist_exp}
      \Pr_{p_E}(\wt(E) \geq d/2) \leq \exp\left(-(d - 2np)^2/2n\right).
\end{align}
For the repetition code with $d=n$, every bad example has weight at least $d/2$, and so we find
\begin{equation}
    \Pr(\text{bad})\leq \exp\left(-\frac{n}{2} (1-2p)^2\right)
\end{equation}
Tighter bounds are available e.g. Ref.~\cite{arratia1989tutorial}. Finally, to show that $\DDD$ is \textit{equivalent} to this classification task we need to also show that the noisy binary classification problem reduces to $\DDD$. We can do so by showing that $EX^{\vec{\eta}}$ (and query access to $f_0$) can be used to construct $D_N$. The following strategy works:
\begin{enumerate}
    \item Sample $(\sigma, \tilde{z}) \sim EX^{\vec{\eta}}$
    \item If $\tilde{z} = 1$, output $(\sigma, f_0(\sigma)):=(\sigma, e)$. Otherwise, output $(\sigma, f_0(\sigma) \oplus 1^n)$
\end{enumerate}
The distribution of $(\sigma, e)$ induced by this procedure is identical to $p_{\Sigma E}$. To see this, first partition the set of syndromes into 
\begin{equation}
    \{\sigma: f^*(\sigma) = f_0(\sigma)\} \cup \{\sigma: f^*(\sigma) \neq f_0(\sigma)\}
\end{equation}
and consider any syndrome  $\sigma$ sampled in step 1 from the first set (i.e. $f^*(\sigma) = f_0(\sigma)$). The joint probability of sampling $(\sigma, \tilde{z}=1)$ is
\begin{align}\label{eq:prz1}
    \Pr_{\sigma, \tilde{g}}(\sigma, \tilde{g}(\sigma) = f^*(\sigma)) = p_\Sigma(\sigma) - \eta(\sigma) = \Pr_{\sigma, \tilde{z}}(\sigma, \tilde{z}=1)
\end{align}
and in this case the output of step 2 above is $(\sigma, f^*(\sigma))$. Meanwhile, the probability of sampling $(\sigma, \tilde{z}=0)$ is
\begin{equation}
    \Pr_{\sigma, \tilde{g}}(\sigma, \tilde{g}(\sigma) \neq f^*(\sigma)) = \eta(\sigma)=\Pr_{\sigma, \tilde{z}}(\sigma, \tilde{z}=0) 
\end{equation}
and in this case the output of step 2 is $(\sigma, f^*(\sigma) \oplus 1^n)$. As these outputs are distributed identically to Eqs.~\ref{eq:ddd_noise0}-\ref{eq:ddd_noise}, the result of this procedure is indistinguishable from sampling from $p_{\Sigma E}$. The same reasoning holds for the set of $\sigma$ for which $f^*(\sigma) \neq f_0(\sigma)$, in which case $\Pr_{\sigma, \tilde{z}}(\sigma, \tilde{z}=1)$ in Eq.~\ref{eq:prz1} is instead $\eta(\sigma)$. Since sampling from $EX^{\vec{\eta}}$ reduces to sampling from $p_{\Sigma E}$ given $f_0$, and we already showed the opposite reduction, these two classification problems are formally equivalent.

\subsection{Proof of Corollary~\ref{cor:1}: Sample complexity of \texorpdfstring{$\DDD$}{DDD} for repetition code}\label{app:cor2}

We can now prove Corollary~\ref{cor:1}, which follows from the reduction from nondegenerate $\DDD$ with $k=1$ to noisy binary classification. This result follows by a slight adjustment to the sample complexity for Probably Approximately Correct (PAC) \cite{valiant1984theory} learning a finite hypothesis class with Random Classification Noise (RCN) \cite{angluin1988learning}. PAC-learnability usually does not apply to instance-dependent noise (IDN), since IDN can be chosen adversarially to corrupt the learning process at a rate higher than the maximum instance noise rate (as in the case of learning half spaces \cite{Massart_2006}). However, we will drop the need for an efficient learning algorithm over a large hypothesis class to bypass this issue.

Let $\mathcal{F}_{n}$ denote the set of Boolean functions on $n$ bits. Consider a (noiseless) examples oracle $EX$ which outputs $(x, c^*(x))$ where $x \in \{0,1\}^n$ is sampled according to $p_X$, and then a label is computed according to some fixed $c^*\in \mathcal{F}_n$. We use this to construct a noisy IDN examples oracle $EX^{\vec{\eta}}$ equipped with a set of instance-dependent noise rates $\{\eta(x): x \in \{0,1\}^n, \eta(x)<p_X(x)/2\}$ which samples $(x, c^*(x)) \sim EX$ and then conditionally outputs\footnote{Note that $\eta(x)$ in this proof plays the role of $\eta(\sigma)/p_\Sigma(\sigma)$ from the previous section.}
\begin{equation}
    (x, \tilde{z}) = \begin{cases}
        (x, c^*(x)), & \Pr=1 - \eta(x) \\
        (x, \neg c^*(x)), & \Pr = \eta(x).
    \end{cases}
\end{equation}
We define $\mathcal{D}:= p_{X\tilde{Z}}$ to be the  distribution induced by sampling under this scheme. For a (noisy) training set $\tilde{D}_N = \{(x_i, \tilde{z}_i)\}_{i=1}^N$ containing $N$ noisy examples sampled iid from $EX^{\vec{\eta}}$, we define
\begin{align}
    L_{\tilde{D}_N}(h):&= \frac{1}{N} \sum_{i=1}^N \I\{h(x_i)\neq \tilde{z}_i\} \\
    L_{\mathcal{D}}(h):&= \Pr_{x, \tilde{z}} (h(x)\neq \tilde{z})
\end{align}
to be the empirical risk and generalization error of a hypothesis $h \in \mathcal{F}_{n}$ corresponding to a 0-1 loss function. Note that $c^*$ minimizes generalization error, achieving the minimum:
\begin{equation}
    \eta:=L_\mathcal{D}(c^*) = \Pr_{x,\tilde{z}}(c^*(x) \neq \tilde{z}) = \sum_{x \in \mathcal{X}}p_X(x) \eta(x).
\end{equation}
We will search $\mathcal{F}_n$ for a function that minimizes empirical risk: $h_{ERM} \in \argmin_{h \in \mathcal{F}_n} L_{\tilde{D}_N}(h)$. We would like to be confident that $L_{\mathcal{D}}(h_{ERM})$ is small under this condition. We will do so by bounding the probability that a ``bad'' hypothesis $h$ (one with large $L_{\mathcal{D}}(h)$) is selected as $h_{ERM}$. Defining 
\begin{align}
    \gamma :&= (1 - 2\max_x\eta(x)),
\end{align}
and computing
\begin{align}
    L_\mathcal{D}(h) &= \sum_{x \in \mathcal{X}} p_X(x) \left((1-\eta(x)) \I\{h(x) \neq c^*(x)\} + \eta(x) \I\{h(x) = c^*(x)\} \right)
    \\&=\sum_{x \in \mathcal{X}} p_X(x) \left((1-\eta(x)) \I\{h(x) \neq c^*(x)\} + \eta(x) (1- \I\{h(x) \neq c^*(x)\})\right)
    \\&=\sum_{x \in \mathcal{X}} p_X(x) \left(\eta(x) + \I\{h(x) \neq c^*(x)\}(1 - 2\eta(x)) \right)
    \\&\geq \eta + \gamma \Pr_x(h(x) \neq c^*(x)),\label{eq:epsbad}
\end{align}
we will call a hypothesis $h$ $\epsilon$-bad if $L_\mathcal{D}(h) > \eta + \gamma \epsilon$. We can show that $L_{\tilde{D}_N}(c^*)$ is close to $L_\mathcal{D}(c^*)$ given sufficiently many samples:
\begin{align}
    \Pr\left(L_{\tilde{D}_N}(c^*) > \eta + \frac{\gamma \epsilon}{2}\right) &= \Pr\left(L_{\tilde{D}_N}(c^*) > L_{\mathcal{D}}(c^*) + \frac{\gamma \epsilon}{2}\right)
    \\&\leq \Pr\left(|L_{\tilde{D}_N}(c^*) - L_{\mathcal{D}}(c^*)| >\frac{\gamma \epsilon}{2}\right)
    \\&\leq 2\exp \left( -\frac{N \gamma^2 \epsilon^2}{2} \right),
\end{align}
where the final inequality follows from applying a Hoeffding bound. Thus, $N = O\left( \frac{1}{(\gamma \epsilon)^2}\right)$ samples are sufficient for $L_{\tilde{D}_N}(c^*)$ to be $\gamma \epsilon/2$-close to $L_{\mathcal{D}}(c^*)$. This event implies that $L_{\tilde{D}_N}(h_{ERM}) < \eta + \gamma \epsilon/2$ since $c^* \in \mathcal{F}_n$. Now we will show that, for the same number of samples the probability that any particular $\epsilon$-bad hypothesis achieves $L_{\tilde{D}_N}(h)< \gamma\epsilon/2$ is very small. From Eq.~\ref{eq:epsbad}, we see that for an $\epsilon$-bad hypothesis, $L_{\tilde{D}_N}(h) < \eta + \gamma\epsilon/2$ implies that $L_{\mathcal{D}}(h) - L_{\tilde{D}_N}(h) > \gamma \epsilon/2$, meaning that
\begin{align}
    \Pr(L_{\tilde{D}_N}(h) < \eta + \gamma\epsilon/2) &\leq \Pr \left(L_{\mathcal{D}}(h) - L_{\tilde{D}_N}(h) > \gamma \epsilon/2\right)
    \\&\leq \exp \left( -\frac{N \gamma^2 \epsilon^2}{2} \right)
\end{align}
Taking a union bound over all $h \in \mathcal{H}$, the probability of an $\epsilon$-bad $h$ being selected as $h_{ERM}$ is less than $\delta$ for a number of samples at least
\begin{align}
    N \geq \frac{2\log (|\mathcal{F}_n|/\delta)}{\epsilon^2 \gamma^2},
\end{align}
which is always satisfied for the choice
\begin{equation}
    N \geq \frac{2\log ( |\mathcal{F}_n|/\delta)}{\epsilon^2 (1 - 2\Pr(\text{bad})^2)}.
\end{equation}
Applying this bound to the distribution induced by $EX^{\vec{\eta}}$ given in the previous Section yields Corollary~\ref{cor:1}. 

\subsection{Class imbalance in \texorpdfstring{$\DDD$}{DDD} versus Proposition~\ref{prop:1}}\label{app:class_imbalance}

Proposition~\ref{prop:1} demonstrated that $\DDD$ for the repetition code is \textit{equivalent to} a noisy binary classification problem. We should therefore justify applying concepts like ``class imbalance'' and ``noise'' directly to $\DDD$ and the results depicted in Fig.~\ref{fig:rep_code_importance}b, even though neither concept is directly relevant to $\DDD$. Here, we answer the question: How do noise and class imbalance in the classification problem posed by Proposition~\ref{prop:1} affect learning the decoding map in $\DDD$?

To connect these tasks, we first assume that $f_0$ is capable of decoding low weight errors. If this were not the case, for example if $f_0(\sigma)$ always returned the highest weight $e$ such that $He = \sigma$, then by Proposition~\ref{prop:1}, the majority of examples would have the label $z(\sigma):=\I\{f_0(\sigma)=f^*(\sigma)\}=0$ (i.e. $f^*(\sigma) \neq f_0(\sigma)$), but it would be counterproductive to reduce this class imbalance. We will assume specifically that $\Pr(f_0(\sigma)=e)$ decreases as the probability of high-weight errors according to $p_E$ increases. Then, 
\begin{enumerate}
    \item In the context of Proposition~\ref{prop:1}, where data are of the form $(\sigma, z(\sigma))$ turning the knob results in a larger probability of examples $(\sigma, 1)$, and therefore a larger probability of syndromes $\sigma$ such that $f^*(\sigma) \neq f_0(\sigma)$. 
    \item Simultaneously, turning the knob increases the probability of high weight errors which, by assumption, increases the probability that $f_0(\sigma) \neq e$. 
\end{enumerate}
These effects combine to increase the probability of syndromes such that $f^*(\sigma) = e$. In the context of $\DDD$ for the repetition code, where data are of the form (syndrome, error), 1.-2. above result in a greater probability of examples $(\sigma, f^*(\sigma))$. Due to the structured nature of $\DDD$, for any fixed $\sigma$ there are only two possible examples: $(\sigma, f^*(\sigma))$ and $(\sigma, \neg f^*(\sigma))$. Since we assume that the NN decoder will try to minimize training loss for each $\sigma$ in the dataset, increasing the probability of the example $(\sigma, f^*(\sigma))$ will decrease the LEP of $\hat{f}$. Therefore, decreasing the class imbalance in Proposition~\ref{prop:1} by turning the knob \textit{combined with the assumption that $f_0$ is effective at decoding low-weight errors} will lead to higher NN decoder performance in the $\DDD$ task.

Similar reasoning applies to label noise in the context of Proposition~\ref{prop:1}: Increasing the label noise for $z(\sigma)$ while simultaneously increasing the probability of high weight errors corresponds to replacing some examples of the form $(\sigma, f^*(\sigma))$ with examples of the form $(\sigma, \neg f^*(\sigma))$ instead. This will tend to increase the LEP of $\hat{f}$ in the $\DDD$ problem.  So, increasing the noise in Proposition~\ref{prop:1} by turning the knob combined with the above assumption for $f_0$ will lead to lower NN decoder performance in the $\DDD$ task.

\subsection{Example importance and well-ordered errors}\label{sec:well_ordered}

We will discuss when and how turning the knob fails for the repetition code experiments discussed in the main text. Recall that in the biased noise model, the first $n/2$ bits have probability $p$ of error, and the second $n/2$ bits have probability $\alpha p$ of error. Recall the minimum weight decoder that works as follows; if we define $\mathcal{E}(\sigma):= \{e: He = \sigma\}$ to be the (pair of) errors whose syndrome is $\sigma$, then:
\begin{equation}\label{eq:minwt_f0}
    f_0(\sigma) = \begin{cases}
        \argmin_{e\in \mathcal{E}(\sigma)}\wt(e) & \text{if }\wt(e) \neq n/2\\
        e \in \mathcal{E}(\sigma) \text{ randomly} & \text{if }\wt(e) = n/2
    \end{cases}
\end{equation}
When is this a good baseline? If the noise model is unbiased ($\alpha=1$), then $f_0$ is $\MLD$ for this code. Conversely, as the bias of the noise model increases, $f_0$ will start to underperform $f^*$. We will say that the set of errors $\{0,1\}^n$ is \textit{well ordered} (with respect to $p_E$) $\wt(e) > \wt(e') \Rightarrow p_E(e) < p_E(e')$ for all $\wt(e) < n/2$. This means that with the exception of weight $n/2$ errors, every low weight error is more likely than every higher weight error. Given this definition, $f_0$ starts to fail with decreasing $\alpha$ or increasing $p$ in one of two ways:
\begin{itemize}
    \item If the errors are well-ordered, but some of the errors with weight $n/2$ are much more likely than others (in which case the guesses of $f_0$ put it at a disadvantage)
    \item If the errors are not well-ordered, then there is at least one error $e$ for which  $f^*(\sigma(e)) = e \oplus 1^n$ while $f_0(\sigma(e)) = e$. Every such error $e$ is an important example with relatively high probability in our dataset.
\end{itemize}
In the toy problem, we can analyze these behaviors precisely. Well-orderedness is violated when either the bias $\alpha$ or the error rate $p$ becomes large: For instance, we can see that the distribution in Fig.~\ref{fig:rep_code_importance} is not well-ordered by the presence of important examples with weight less than $n/2$, whereas $p=0.1, \alpha=0.7$ imply that the distribution $p_E[p, \alpha]$ is well-ordered. 

However, the training distribution $p_E[\beta p, \alpha]$ resulting from turning the knob is no longer well-ordered for sufficiently large $\beta$. This transition point occurs at the combination of $(\alpha, p, \beta)$ such that the highest-probability error with weight $n/2+1$ in $p_E[\beta p, \alpha]$ is more likely than the lowest-probability error with weight $n/2-1$ in the same distribution. For $n=8$, and denoting a $x$-bit as a bit in this error model that has probability $x$ of undergoing bitflip, then most likely weight-5 error involves all four $\beta p$-bits flipping along with one $\alpha \beta p$-bit flipping, which occurs with probability $(\beta p)^4 (\alpha \beta p) (1-\alpha \beta p)^3$. The least likely weight-3 error involves three $\alpha \beta p$-bits flipping, which occurs with probability $(\alpha \beta p)^3 (1-\alpha \beta p) (1-\beta p)^4$. The violation of well-orderedness therefore occurs for
\begin{equation}
    \frac{(\beta p)^4}{(1-\beta p)^4} > \frac{(\alpha \beta p)^2}{(1 - \alpha \beta p)^2}.
\end{equation}
Once $\beta$ is sufficiently large to satisfy this inequality, the $\MLD$ for $p_E[\beta p, \alpha]$ is substantially different than the $\MLD$ for $p_E[p, \alpha]$, e.g. results in an error rate  $O(p^3)$ higher. Consequently, the NN decoder will maximize its accuracy with respect $p_E[\beta p, \alpha]$ by adopting a strategy that is guaranteed to be wrong when the model is evaluated on $p_E[p, \alpha]$.

Thus, one of the interesting aspects of turning the knob for this toy problem is the existence of a discrete transition in behavior due to the violation of well-orderedness. Another salient feature is that we can efficiently compute the sample importance among well-ordered distributions, for example showing that good and bad examples are nearly equiprobable. To this end, assume that $p_E[p, \alpha]$ is well-ordered, i.e. 
\begin{equation}
    \frac{(p)^4}{(1-p)^4} < \frac{(\alpha p)^2}{(1 - \alpha p)^2}.
\end{equation}
Denote by $W(n, i)$ as the set of bitstrings $x \in \{0,1\}^n$ with $\wt(x) = i$. For the toy problem, we compute each value for even $n$. Exactly half of the elements of $W(n, n/2)$ are good examples, which are enumerated in nonincreasing order by how many bits are set on the left half starting from $1^{n/2} 0^{n/2}$. Then, for each $x \in W(n, n/2)$ the baseline $f_0$ randomly guesses whether to decode $x$ or $x \oplus 1^n$ and is correct half of the time, so that half of the aforementioned good examples are important. For even $n$ let us define
\begin{align}
    f_n(x, y) &= \sum_{k=0}^{\ceil{n/4}-1} \binom{n/2 }{k}^2  x^{n/2-k} (1-x)^k y^k (1-y)^{n/2 - k} + R_n  \\
    R_n &=\begin{cases}
        \frac{1}{2} \binom{n/2}{n/4}^2  x^{n/4} (1-x)^{n/4} y^{n/4} (1-y)^{n/4} & {n\mod 4} = 0 \\
        0 & \text{else}
    \end{cases}\label{eq:fnxy}
\end{align}
Then we have that 
\begin{align}
    \Pr_{p_E[p, \alpha]} (\text{important examples}) = f_n(p, \alpha p)
\end{align}
Then we can see that the remainder term $R_n$ in Eq.~\ref{eq:fnxy} exists because whenever $4$ evenly divides $n$ we have to divide evenly the examples having $n/4$ ones on the left and right side into good and bad. The probability of bad examples is just the probability of the set of complements of good examples, given by $f_n(p_2, p_1)$ plus the set of all bitstrings having weight strictly greater than $n/2$:
\begin{align}
    \Pr_{p_E[p, \alpha]} (&\text{bad examples}) = f_n(\alpha p, p) \\&+ \sum_{w=n/2+1}^n \,\,\sum_{\substack{0 < k,\ell \leq n/2:\\ k+\ell = w}} \binom{n/2}{k}\binom{n/2 }{w-k} p^k (1-p)^{n/2-k} (\alpha p)^{w-k} (1-\alpha p)^{n/2-(w-k)}
\end{align}
From this, we see that without turning the knob, for well-ordered distributions (i.e. bitflip models with low bias), there is roughly the same amount of signal and noise. This provides more evidence that turning the knob is a valuable tool for overcoming the effect of bad examples on $\DDD$

\section{Learning theory for \texorpdfstring{$\DDDD$}{Detector DDD}}

\subsection{Proof of Proposition~\ref{prop:2}}

The proof of Proposition~\ref{prop:2} is similar to proof of Proposition~\ref{prop:1}. However, we first need change how we define an $\MLD$ since data in $\DDDD$ only provide information on one component ($\bar{X}$ or $\bar{Z}$) of a logical error. Without loss of generality, consider a $\DDDD$ dataset $D_N: \{(\sigma_i, y_i)\}$ (where $\sigma_i$ represents all stabilizer measurements overall timesteps $t=0,\dots, T$) and a random variable $Y = \I\{ \langle \bar{Z}\rangle^{(t=T+1)} \neq \langle \bar{Z}\rangle^{(t=0)}\}$. Then, the maximum-likelihood label for this problem is defined as  
\begin{equation}
    y^*(\sigma) =\argmax_{y \in \{0,1\}} \Pr_{p_E}(Y=y|\Sigma=\sigma).
\end{equation}
This maximum likelihood label $y^*(\sigma)$ for a syndrome $\sigma$ can be computed using an $\MLD$ $f^*$ by first computing the most likely logical error $ \bar{E} = f^*(\sigma)$ and then outputting ``1'' if $\bar{E} \in \{\bar{X}, \bar{Y}\}$. Thus, if a logical error $\bar{E}_i$ gives rise to the syndrome $\sigma_i$, then $f^*(\sigma_i) = \bar{E}_i$ implies that $y^*(\sigma_i) = y_i$. We can similarly define a baseline label $y_0: \Sigma \rightarrow \{0,1\}$ as the $\bar{X}$-component of a baseline decoder $f_0$. Good, bad, and important examples are now defined with respect to $y_0$ and $y^*$, e.g. $\Pr(\text{good}) = \Pr_{(\sigma, y)}(y = y^*(\sigma))$.

We now re-label samples according to $z(\sigma) := \I \{y_0 (\sigma) = y^*(\sigma)\}$, and then sample a noisy dataset of pairs $(\sigma, \tilde{z}(\sigma))$ according to the procedure
\begin{enumerate}
    \item Sample $ \sigma \sim p_{\Sigma}$
    \item Sample a noisy optimal label $\tilde{y}$  that outputs
    \begin{equation}
        \tilde{y}(\sigma) = \begin{cases}
            y^*(\sigma)& \Pr=1 - \eta(\sigma)/p_\Sigma(\sigma) \\
            y^*(\sigma) \oplus 1, & \Pr = \eta(\sigma)/p_\Sigma(\sigma)
        \end{cases}
    \end{equation}
    \item output $(\sigma, \tilde{z})$ with label $\tilde{z}(\sigma) =\I\{y_0(\sigma) = \tilde{y}(\sigma)\}$
\end{enumerate}
We can similarly compute $\Pr_{\sigma, \tilde{z}}(\tilde{z}(\sigma) \neq z(\sigma)\} = \Pr(\text{bad})$, and show that class priors obey
\begin{align}
    |\Pr_{\sigma}(z(\sigma) = 0) - \Pr(\text{important})| &\leq \Pr(\text{bad})\\
    |\Pr_{\sigma}(z(\sigma) = 1) - \Pr(\text{unimportant})| &\leq \Pr(\text{bad})
\end{align}
Finally, note that if a logical error $\bar{E}$ induces a logical error flip $y$, then $y^*(\sigma) \neq y$ implies $f^*(\sigma) \neq \bar{E}$. If we assume an iid depolarizing model with probability $(1-p)$ of no error, we can compute
\begin{align}
    \Pr(\text{bad}) &= \Pr_{(\sigma, y)}(y^*(\sigma) \neq y) 
    \\&\leq  \Pr_{\bar{E}, \sigma}(y^*(\sigma) \neq \bar{E}) 
    \\&\leq \Pr_{p_E}\left(\wt(E) \geq \frac{d}{2}\right)
    \\&\leq \exp\left(-(d - 2np)^2/2n\right).
\end{align}
where we have used Eq.~\ref{eq:dist_exp} in the final line.

\subsection{Alignment}

When training a model via turning the knob, we can bound the distance of the model to the correct $\MLD$ in terms of training error on the high-$\beta$ distribution and misalignment between two distributions.

First, observe that for any function $g: \Sigma\rightarrow \{0,1\}$ acting on the set of syndromes, and two distributions $p,q: \mathcal{\Sigma}\rightarrow \mathbb{R}^+$, we have
\begin{align}
    \E_{\sigma\sim p}[g(\sigma)] - \E_{\sigma\sim q} [g(\sigma)] &= \sum_{x \in \Sigma} g(\sigma)[p(\sigma) - q(\sigma)] 
    \\&\leq \sum_{\sigma \in \Sigma} |p(\sigma) - q(\sigma)| 
    \\&= \norm{p - q}_1 \label{eq:tvbound}
\end{align}
Now let $f^*_p, f^*_q$ be $\MLD$ decoders for distributions $p$ and $q$ respectively. Applying Eq.~\ref{eq:tvbound} with $g(\sigma) := \I\{h(\sigma) \neq f_q^*(\sigma)\}$, we find that for any decoder $h$
\begin{align}
    \Pr_{\sigma \sim p}\left(h(\sigma) \neq f^*_p(\sigma)\right) &\leq \Pr_{\sigma \sim p}(f_p^*(\sigma) \neq f_q^*(\sigma)) + \E_{\sigma \sim p} \left( \I\{h(\sigma) \neq f_q^*(\sigma)\}\right)
    \\&\leq \Pr_{\sigma \sim p}(f_p^*(\sigma) \neq f_q^*(\sigma) )+ \E_{\sigma \sim q} \left( \I\{h(\sigma) \neq f_q^*(\sigma)\}\right) + \norm{p - q}_1
    \\&=  \Pr_{\sigma \sim p}(f_p^*(\sigma) \neq f_q^*(\sigma) ) + \Pr_{\sigma \sim q}\left(  h(\sigma) \neq f_q^*(\sigma) \right) + \norm{p - q}_1
\end{align}

\section{Experiments}\label{sec:experiments}
Here we provide detailed descriptions of the machine learning experiments summarized in the main text. As we used fundamentally different methods for experiments in $\DDD$ versus $\DDDD$, Sec.~\ref{app:dddexp} refers exclusively to the repetition code and single-round surface code experiments, while Sec.~\ref{sec:DDDD_exp} refers exclusively to multi-round surface code experiments unless otherwise noted.

\subsection{\texorpdfstring{$\DDD$}{DDD} Experimental methods}\label{app:dddexp}

\textbf{Baseline decoders} We focus on the following baseline decoders:
\begin{enumerate}
    \item Naive lookup table: The lookup table $f_{D_N}$ will, given an input syndrome $\sigma$, output the error $e$ that appears with highest frequency among pairs $(\sigma, e)$ in $D_N$. If $\sigma$ does not appear in any example in $D_N$, $f_{D_N}(\sigma) = I$. The lookup table \textit{does not} guess among the two $e$ satisfying $\sigma = He$, as this would encode domain knowledge for the coding problem. Another strategy for dealing with unseen $\sigma$ is to guess among all $e \in \{0,1\}^n$; this would not change the accuracy in any important way.
    \item Minimum weight decoder: For the toy problem, we used the standard ``minimum weight decoder'' for the repetition code, which outputs the lowest-weight error such that $He=\sigma$. For the surface code experiment, we used Minimum Weight Perfect Matching with edge weights equal to $\log((1-p_i)/p_i)$ for each qubit with depolarizing probability $p_i$.
    \item Minimum Weight Perfect Matching: For the surface code experiments, we consider  MWPM with a matching graph weighted according to $\log( (1-p_i)/p_i)$ as a baseline. While there exist variations of MWPM better suited for correlated errors or incorporating calibration data \cite{google2023suppressing,sivak2024optimizationdecoderpriorsaccurate}, we do not consider those here as we are mainly interested in the relative improvement to $\DDD$ by turning the knob. 
\end{enumerate}

\textbf{Efficient virtual training} We only use weighted loss and accuracy functions for training and evaluation, which speeds up training runs by several orders of magnitude compared to actually sampling data. $\DDD$ is an algorithmic learning task where $N$ training examples $(x, y)$ having a functional relationship (either $x = x(y)$ or $y=y(x)$) are sampled from a discrete set $\mX \times \mY$. However, it is (usually) unnecessary to have $N > |\mX|$ due to the pigeonhole principle. Instead, a loss function should just be reweighted according to a histogram of $N$ samples from $p_X$ distributed among $|\mX|$ bins -- or, in the case of mini-batching with batch size $m$, there should be $N/m$ such histograms each containing $m$ samples. In classical decoding, the syndrome $\sigma = \sigma(e)$ is a function of the error $e$. If $n_i(e)$ is the number of times the error $e$ appears in batch $i$, then the \textit{virtual training loss} for batch $i$ is defined as
\begin{align}
    L(f) &= \sum_{\substack{e \in \{0,1\}^n: \\ n_i(e) \neq 0}} \ell(f(\sigma(e)), e) \frac{n_i(e)}{m}\label{eq:virtual_tr}
    \\&= \sum_{k=1}^m \ell(f(\sigma(e_k)), e_k)
\end{align}
where the final line is the ordinary batched loss, and $\{e_1, \dots, e_m\}$ is a set of errors sampled iid from $p_E$ to compose the batch. The complexity of batched and unbatched training are therefore equivalent whenever $m > 2^n$. Since $p_E$ typically concentrates on a few errors $e$, the vector of batch counts $[n_i(e^{(1)}), \dots, n_i(e^{(2^n)})]$ is stored as a ragged array with length equal to the number of nonzero $n_i$ sampled for that batch. In this way, the computational cost per training step (i.e. single gradient computation) is typically equivalent to computing ordinary training loss for a few dozen samples, yielding several orders of magnitude speedup in training time. Computation of validation loss can be similarly sped up. While the per-epoch training time could be sped up significantly by setting batchsize equal to $N$, we  use minibatches so as to preserve the training dynamics of multiple gradient updates per epoch, as is common practice in real training scenarios. Furthermore, virtual training allows us to simulate training on an \textit{infinite amount of data} as shown in Fig.~\ref{fig:toric_results}, which is accomplished by replacing $n_i(e)/m$ with $p_E(e)$ in Eq.~\ref{eq:virtual_tr}. Training accuracy and validation accuracy are computed as
\begin{equation}
    \sum_{e \in \{0,1\}^n} \I\{f(\sigma(e)) = e\} p_E(e) 
\end{equation}
and have computational cost equal to computing accuracy on $2^n$ samples. For our toy problem, $2^8 = 256$ is much smaller than the training data size, and so again the accuracy computations are orders of magnitude faster. The accuracy computation is a bottleneck at roughly $\sim 10 \times$ the cost of the loss computation, so we evaluate accuracy fewer than once per epoch (usually once per 10 epochs). These speedups for accuracy and loss computation allow us to run hundreds of times more models than with ordinary training methods. 

\textbf{Verification and hyperparameter pruning} Many of our experiments are designed probe how the long-tailed $p_E$ affects training dynamics. To do so reliably we must control for the algorithmic complexity of $\DDD$, i.e. demonstrate that a model that fails to learn from $p_E$ does so due to the sparsity of important examples, rather than a lack of capacity to solve the decoding problem in general. We therefore only present results from models that have been verified capable of achieving perfect accuracy on the given task. To perform verification, we used a modified distribution of errors $p_E'$ that is uniform over the good examples (with respect to the original $p_E$), e.g. $p_E(e) = 2^{-(n-1)}\cdot \I(e \text{ is good})$ for the classical repetition code. We then performed hyperparameter search to identify architectures that were capable of perfectly solving $\DDD$ and learning $\MLD$ under the ideal distribution $p_E'$. We used this process to adjust hyperparameter ranges so that at least roughly 70\% of models were capable of achieving optimal performance given uniform good examples, and discarded hyperparameters that consistently resulted in failed runs.

\textbf{Architectures}  We used the following architectures for the $\DDD$ experiments with turning the knob. For the toy problem we used three architectures, while we only used transformers for the $[[9, 1, 3]]$ rotated surface code $\DDD$ experiments. Hyperparameters for these architectures follow in Tables~\ref{tab:hypers} and~\ref{tab:hypers2}.

\begin{itemize}
    \item \textit{Feedforward Neural Network (FNN)}. We implemented a standard multilayer perceptron with $n_{layers}$ hidden layers each having the same hidden dimension of $\text{width}$.
    \item \textit{Convolutional Neural Network (CNN)}. For the toy problem, the repetition code syndromes do not contain any 2D geometrical information, so we use 1D convolutions with $n_{conv}$ convolution channels followed by a final linear layer with 8 dimensional output. 
    \item \textit{Encoder-Decoder transformer}. We use the basic encoder-decoder transformer architecture of \cite{vaswani2017attention} with $n_{enc}$ encoder layers and $n_{dec}$ decoder layers. We do not tokenize inputs. Since we generate fixed-length outputs autoregressively, we reuse the $0$ bit for start-of-sequence/end-of-sequence tokens without any impact on performance. Self-attention and cross-attention modules each had $h$ heads, model embedding dimension was $d_{model}$, elementwise feedforward network dimension was $d_{FFN}$.
\end{itemize}

\textbf{Noise model} For the repetition code experiments, we used a symmetric bitflip model with $n=8$ bits, the first four have probability $0.1$ of bitflip, the last four have probability $0.07$ of bitflip. For the repetition code experiments, we applied $n=9$ uncorrelated, symmetric depolarizing channels each with probability $(1-p_i)$ of no error for $i=1\dots 9$. The set $\{p_i\}$ was sampled from a normal distribution with mean $\bar{p}=0.05$ and variance $\text{var}=0.03$.

\textbf{Hyperparameter tuning} For the toy problem experiments in Fig.~\ref{fig:rep_code_importance}, we randomly sampled hyperparameters from intervals/sets in Table~\ref{tab:hypers} (red parameters denote hyperparameters that were ruled out during the verification phase and were not sampled to generate Fig~\ref{fig:rep_code_importance}). Roughly 150-200 models were trained per choice of $p_{train}$. We followed a similar procedure to produce Fig.~\ref{fig:toric_results} using hyperparameters in Tab.~\ref{tab:hypers2}. In both cases, we selected the range of hyperparameters by first training a a few hundred models on only good examples and verifying that roughly $\geq 70\%$ of models could learn the corresponding $\MLD$ decoding rule.

\begin{table}
    \centering
    \begin{tabular}{lll}
           Transformer &  CNN & FNN \\ 
           \hline
           $\text{lr}\in \{  10^{-5},\dots ,   2\times 10^{-3}
         \}$  &  $\text{lr}\in \{ 1\times 10^{-5}, \dots, 2\times 10^{-3}\}$  & $\text{lr}\in\{ 3\times 10^{-4},\dots, 10^{-2}\}
         $  \\
        $n_{enc} \in \{1, 2, 3\}$  & $n_{layers} \in \{\red{2}, 3, \dots, 6\}$  & $n_{layers} \in \{2, 3, 4, \red{5, 6}\}$ \\
          $n_{dec} \in\{1, \red{2, 3}\}$ & kernel size$\in\{\red{1, 2}, 3, 4\}$ & $\text{width}\in\{\red{8, 16}, 32, 64\}$    \\
          $h \in \{2, 4\}$ & $n_{conv}\in\{\red{1, 2, 3, 8}, 16, 24, 32\}$  & \\
          $d_{model}\in\{8, 12, 16\}$ & & \\
          $d_{FFN} \in \{4, 8, 12\}$ &  &
    \end{tabular}
    \caption{\textbf{Hyperparameter ranges for $\DDD$ toy problem.} All models were optimized with adam optimizer using the given learning rate. (Virtual) batch sizes between $\{100, \dots, 500\}$ did not affect performance, final batch size was 100 (FNN, CNN or 250 (Transformer). Numbers in \red{red} were pruned from the hyperparameter pool due to consistently bad performance on datasets containing only good examples (see Verification); the remaining hyperparameters result in a model capable of learning $\MLD$ at least (roughly) $70\%$ of the time. All models were trained with a weight dropout sampled from $\{5\%, 10\%, 15\%\}$.}
    \label{tab:hypers}
\end{table}
\begin{table}
    \centering
    \begin{tabular}{l}
            \hline
           learning rate $\in${\small $ \{ \red{1\times 10^{-4}}, 5\times 10^{-4}, 8\times 10^{-4}, 1\times 10^{-3}, 2\times 10^{-3}, 3\times 10^{-3}, \red{5\times 10^{-3}}\}$}   \\
           dropout$\in\{0.1, 0.15\}$   \\ 
            $n_{enc} \in \{1, 2, 3\}$   \\
          $n_{dec}=1$      \\
          $h \in \{2, 4\}$    \\
          $d_{model}\in\{\red{4, 8}, 16, 20, 24\}$  \\
          $d_{FFN} \in \{4, 8, \red{16}\}$  
    \end{tabular}
    \caption{\textbf{Transformer hyperparameter ranges for $[[9, 1, 3]]$ surface code $\DDD$ experiments.} Batch size was 256. See caption of Table~\ref{tab:hypers} for other details.}
    \label{tab:hypers2}
\end{table}

\textbf{Training and model selection} Models were trained for up to 10000 epochs. We report the model performance at the epoch with minimum validation error (i.e. ``early stopping''); this corresponds to early stopping according to a maximizing cross-validation with a sufficiently large test set. This allows us to compare the performance of models with different $p_{train}$ under \textit{idealized conditions} where models can be selected according to their test performance. Models were trained in parallel on Intel Xeon Gold 6230 CPUs. For the toy problem, each CNN could be trained in $\sim5$ minutes and each transformer in $\sim30$ minutes, resulting in $\sim 900$ CPU hours (wall time) to compute Fig.~\ref{fig:rep_code_importance}. For the surface code, each transformer took $\sim 60$-$90$ minutes to train to completion consuming $\sim 5000$ CPU hours (wall time) for Fig.~\ref{fig:toric_results}. In Fig.~\ref{fig:toric_results}b of the surface code experiments, we chose $N=10000$ since this is $\approx 4\times$ corresponding to a discrete problem space of $(\sigma, y)$ pairs that is $4\times$ larger. Fig~\ref{fig:toric_results}c shows that this is also the largest $N$ for which $\beta=1.75$ significantly outperforms $\beta=1$.

\begin{figure}[ht]
    \centering
    \includegraphics[width=.8\linewidth]{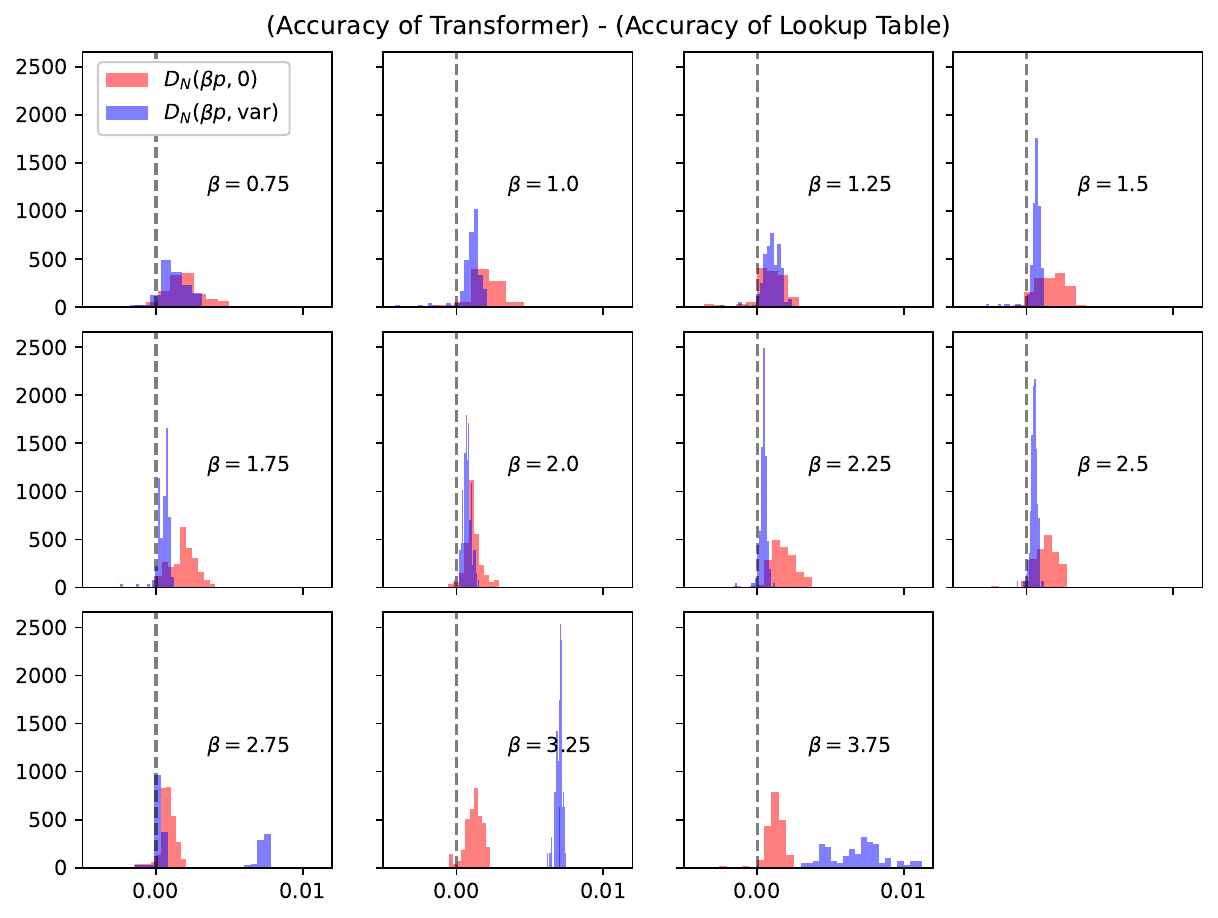}
    \caption{Transformers trained for $\DDD$ on the surface code consistently outperform memorization, demonstrating generalization from training data. This is especially pronounced for large $\beta$, past the threshold for which $\MLD$ for $p_E[\beta p, \text{var}]$ is no longer $\MLD$ for $p_E$.}
    \label{fig:toric_vs_lookup}
\end{figure}

\subsubsection{Additional experiments and results}\label{app:additional}
Fig.~\ref{fig:toric_vs_lookup} is a companion plot to Fig.~\ref{fig:toric_results} in the main text. It shows that majority of transformers trained for the surface code $\DDD$ problem outperformed memorization on the same training data. 

\FloatBarrier

\subsection{Turning the knob with and without device knowledge}\label{app:toric_extra}

\begin{figure}[ht]
    \centering
    \includegraphics[width=\linewidth]{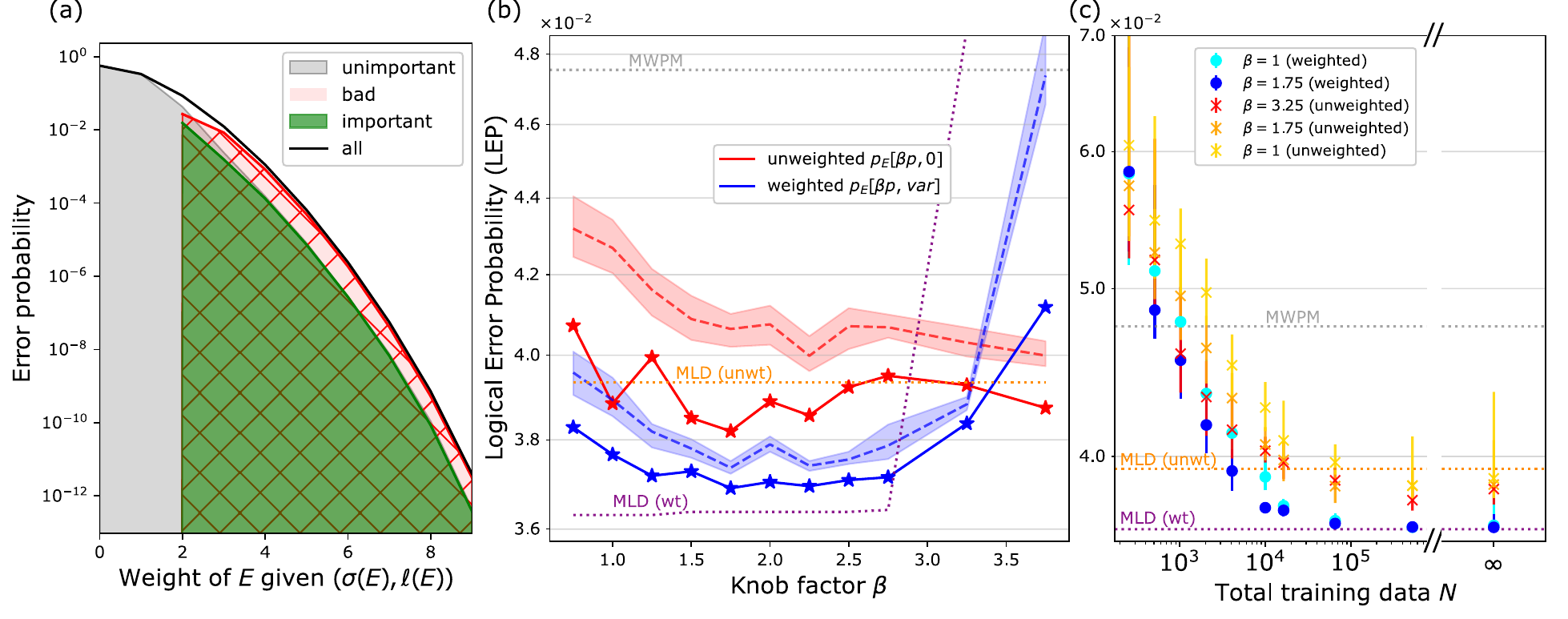}
    \caption{ \textbf{(a) Important examples in the $d=3$ surface code are mixed in with bad examples}, making it harder to predict the benefits of data augmentation. The green area shows total importance of each set of weight-$w$ errors, while the red area shows the cumulative probability of bad examples. \textbf{(b) Turning the knob improves NN decoders for a QECC when the underlying error distribution is partially known.} For each $\beta$ we trained transformers on $N=10^4$ examples sampled from either the unweighted distribution \textcolor{red}{$p_E[\beta p,0]$} that is independent of $p_E$ (strategy 1) or the weighted distribution \textcolor{blue}{$p_E[\beta p,\sigma^2]$} that incorporates prior knowledge about $p_E$ (strategy 2). For strategy 2, increasing $\beta$ improved NN decoder accuracy compared to $\beta=1$. Strategy 1 was on average not helpful, though the strategy 1 models trained with large $\beta$ are competitive with training on $D_N$. \textcolor{orange}{MLD (unwt)} denotes the error of $\MLD$ for $p_E[\beta p,0]$ evaluated on $p_E[p,\sigma^2]$, while \textcolor{purple}{MLD (wt)} denotes the error of $\MLD$ for $p_E[\beta p,\sigma^2]$ evaluated on $p_E[p,\sigma^2]$. Shading denotes IQ range around the median (dashed line) of 150 transformers per $\beta$ ($\sim 3000$ NN decoders total) trained with different initializations and hyperparameters. Solid line with $\star$ indicates best-of-150 accuracy was always superior to a lookup table (additional plots in Appendix~\ref{app:additional}). \textbf{(c) For smaller datasets, turning the knob without prior knowledge of $p_E$ can improve decoder performance,} but as $N\rightarrow \infty$ strategy 1 approaches the suboptimal \textcolor{orange}{MLD (unwt)}.}
    \label{fig:toric_results}
\end{figure}

We again study example importance in the $[[9, 1, 3]]$ rotated surface code and show that turning the knob remains an effective strategy for improving NN decoders. To expedite experiments, we use much larger error rates than the main text, using $p=0.05$ and variance $\sigma^2=0.03$. This has the effect of decreasing both the improvement in LEP due to to turning the knob (baseline error rates are already high) but decreases the range of potentially good $\beta$ values.

Fig.~\ref{fig:toric_results}a shows the distribution of important and bad examples for the surface code with respect to a MWPM baseline decoder, showing significant differences to the classical case in Fig.~\ref{fig:rep_code_importance}: (i) there are important errors for every error weight greater than $1$ due to the variety of stabilizer weights, though the example importance predictably peaks at $\lceil{d/2}\rceil$ and then decays for higher error weights, and (ii) there is a relatively larger probability of bad examples compared to important examples, due to decoding degeneracy: If $\ell, \ell'$ are both $\MLD$ outputs, then only one of $(\sigma, \ell)$ or $(\sigma, \ell')$ can be important and the other is bad.

Similarly to Sec.~\ref{sec:qDDD} we train NN decoders for $\DDD$ using a dataset sampled from a distribution with higher error rates than the true distribution $p_E:=p_E[p, \sigma^2]$. Here, we are interested in how prior knowledge of the error model affects turning the knob, we consider two strategies: 
\begin{itemize}
    \item \textbf{strategy 1}, we train on a dataset $D_N[\beta p, 0]$ containing $N$ examples sampled from an \textit{unweighted} error distribution $p_E[\beta p, 0]$ that encodes no underlying knowledge of $p_E$
    \item \textbf{strategy 2}, we train on a dataset $D_N[\beta p, \sigma^2]$ of $N$ errors sampled from a \textit{weighted} distribution $p_E[\beta p, \sigma^2]$ corresponding to independent depolarizing error probabilities $\{\beta p_i\}$, which encodes the biases of $p_E$.
\end{itemize}
Thus, strategy 1 increases error rates generally, while strategy 2 increases error rates but results in an error distribution qualitatively similar to $p_E$. Note that $D_N[\beta p, \sigma^2]$ technically corresponds to samples from a distribution with mean error $\beta p_i$ and variance $\beta^2 \sigma^2$, since the variance of the distribution is ultimately controlled by increasing individual error rates. 

Fig.~\ref{fig:toric_results}b shows the result of strategies 1-2, confirming that incorporating prior knowledge of error rates makes turning the knob more effective. For larger $\beta$, NN decoders are more accurate when trained on $D_N[\beta p, \sigma^2]$ (strategy 2) versus $D_N[p, \sigma^2]$. For fixed $N$, training on $D_N[\beta p, 0]$ (strategy 1) does not reliably improve model accuracy, which trends towards $\MLD$ for $p_E[\beta p, 0]$. For strategy 2, the accuracy of NN decoders trained on $D_N[\beta p, \sigma^2]$ again exhibits a ``U'' shape. While the LEP initially decreases with $\beta$, for a large enough $\beta$ there will be some syndrome $\sigma$ such that the most likely logical error given $\sigma$ with respect to $p_E[\beta p, \sigma^2]$ differs from the $\MLD$ logical error with respect to $p_E$. The NN decoders will try to learn the $\MLD$ given $D_N[\beta p, \sigma^2]$, and they see a significant drop in accuracy coinciding with the rapidly increasing error of $\MLD$ for $p_E[\beta p, \sigma^2]$ in Fig.~\ref{fig:toric_results}. Fig~\ref{fig:toric_results}c demonstrates that training on $D_{N'}[\beta p, 0]$ can lead to more accurate models when $N$ is small and $N' > N$, but strategy 1 accuracies only ever match $\MLD$ for $p_E[\beta p, 0]$ (or slightly exceeds $\MLD$, due to cross-validation on $D_N$).

\subsection{\texorpdfstring{$\DDDD$}{Detector DDD} Experimental methods}\label{sec:DDDD_exp}
We provide additional details for the experiments in Sec.~\ref{sec:detectorddd} and complementary results to Fig.~\ref{fig:gnn_results_val}.

\begin{table}[h]
    \centering
    \begin{tabular}{l}
           learning rate {\small $\in \{ \red{10^{-4}}, 2 \times 10^{-4}, 3\times 10^{-4}, 4 \times 10^{-4}\}$}   \\
            $n_{conv} \in \{\red{3}, 4,5, 6, \red{7}\}$   \\
          $d_{conv} \in \{\red{16}, 32, 64\}$   \\
          $d_{mlp} \in \{\red{32}, 128, \red{256}\}$      \\
          $n_{mlp} \in \{3, 4\}$    \\
    \end{tabular}
    \caption{\textbf{GNN  hyperparameter ranges for $[[9, 1, 3]]$, $T=5$ rotated surface code $\DDDD$ experiments.} We use the architecture from Ref.~\cite{lange_data_driven_2023} in all experiments. Batch size was either 512 or 1024, with little effect on final accuracy. Hyperparameters were pruned based on which consistently resulted in models with low validation error, and were discarded if they greatly increased parameter count without improving accuracy. Overall, model performance was mostly insensitive to hyperparameter choice within the settled range. Not shown: We tried batch size 128 with learning rates about $10\times$ larger and found a similar range of ``good'' hyperparameters.}
    \label{tab:hypers3}
\end{table}

\begin{figure}[h]
    \centering
    \includegraphics[width=\linewidth]{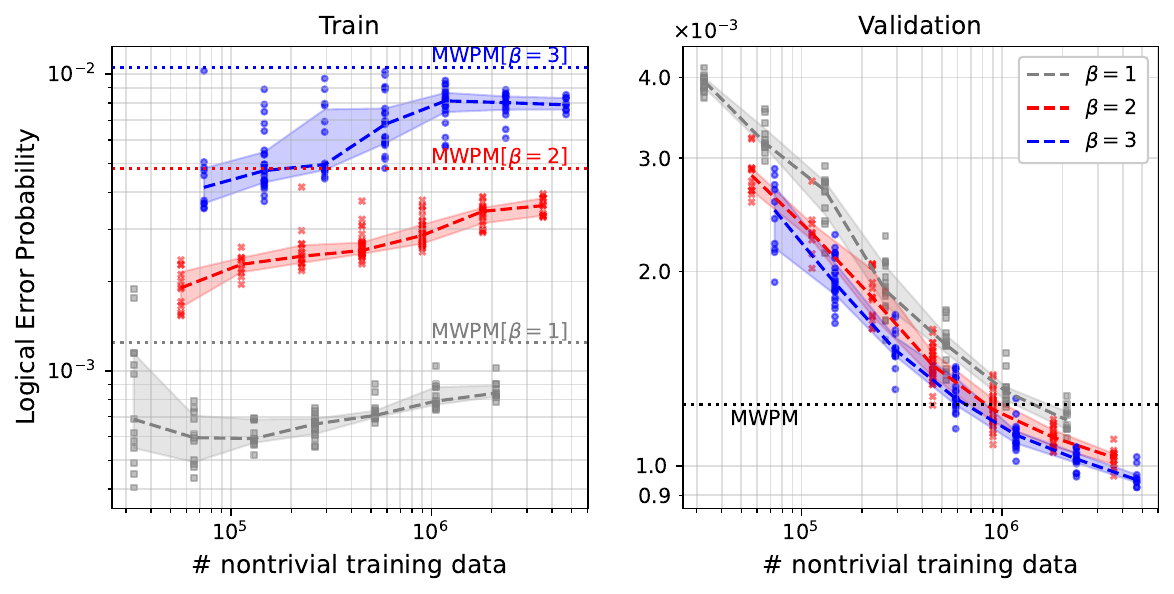}
    \caption{\textbf{Turning the knob still improves training and validation accuracy even when controlled for the number of nontrivial training data still show .} (left) Training error (and corresponding MWPM performance) for $p_E[\beta p]$ with $\beta=1,2,3$ and (right) validation error for the GNN experiments, plotted against the number of nontrivial training data. Since $p \ll 1$, many training examples in $D_N[\beta p]$ will correspond to no error having occurred, resulting in a trivial example trivial to decode. A first-order effect of increasing $\beta$ is to introduce more nontrivial examples into the dataset $D_N[\beta p]$, which effectively increases the amount of training data. Even after controlling for this effect, validation accuracy improves with $\beta$, suggesting that sample importance provides a more robust explanation for the performance improvements.}
    \label{fig:gnn_nontrivial}
\end{figure}

\textbf{Baseline and noise model} Training and validation data were generated by simulating multiple rounds of $Z$-stabilizer measurements for a rotated surface code memory experiment with circuit-level depolarizing noise at rate $p$ before each round of measurement and after each Clifford gate, as well as bitflip rate $p$ before each measurement, as implemented in stim \cite{Gidney_2021}. Fig.~\ref{fig:gnn_results_val} and Fig.~\ref{fig:gnn_nontrivial} show the accuracy for MWPM computed using pymatching \cite{higgott2023sparse} on $10^8$ repetitions of the experiment.

\textbf{GNN architecture and hyperparameters} We used the Graph Neural Network (GNN) architecture developed in Ref.~\cite{lange_data_driven_2023} for $\DDDD$ experiments and turning the knob. Each GNN consists of a sequence of graph convolution (GraphConv) layers \cite{welling, gcn_ref} followed by a multi-layer perceptron (MLP) and then a global mean pooling layer. In the interest of of demonstrating \textit{robust} performance improvements due to turning the knob, we introduced hyperparameters to control the depth and width of both GraphConv and MLP modules. We use a parameters $n_{conv}, d_{conv}$ to denote the number and maximum width of the GraphConv layers. The sequence of layer dimensions follows an expanding-then-contracting pattern $(d_{conv}, 2 d_{conv}, \dots, 2^{n_{conv}/2} d_{conv}, \dots, d_{conv})$. $n_{mlp}, d_{mlp}$ describe the number of MLP layers and maximum MLP hidden dimension. The MLP layer dimensions contract in the sequence $(d_{mlp}, 2^{-1}d_{mlp}, \dots)$.

\textbf{Data structure} Data for the $\DDDD$ problem are more complex than for $\DDD$. We preprocess data according to Ref~\cite{lange_data_driven_2023}: we first construct an $(x, y, t)$ coordinate system for stabilizers where $x,y$ denotes spatial position in the rotated surface code and $t$ denotes the measurement round. Each coordinate is concatenated with a one-hot encoding of whether the corresponding stabilizer is $X$-type or $Z$-type. The 5-dimensional vectors are then used to annotate an graph whose vertices correspond to stabilizers, and the graph is assigned a label $y$ according to whether the observable value $\langle Z_L\rangle$ flipped between $t=0$ and $t=T$. The graph is all-to-all connected with edge weights given by inverse euclidian distance in $(x, y, t)$-space.

\textbf{Training and validation} For each training run we sampled $N\in\{2^{17}, 2^{18}, \dots, 2^{23}\}$ data and trained for up to 3000 epochs. We computed validation accuracy against the true error distribution $p_E$ every 10 epochs, with early stopping after 100 epochs of no improvement of validation score. We speculate that this choice of training on $p_E[\beta p]$ and performing cross-validation with respect to $p_E[p]$ is responsible for our consistent performance improvements, in contrast with models reported in Table 1 of Ref.~\cite{chamberlandTechniquesCombiningFast2023} which may have only been trained and validated with respect to $p_E[\beta p]$. Training time was $\sim 2$ hours for the smallest parameter count/training dataset size/knob factor, $(N=2^{17} \approx 1.3 \times 10^5, \beta=1)$, to $\approx 60$ hours for the largest of these ($N=2^{23}\approx 8.4\times 10^{6}$, $\beta=3$). Total CPU time for Fig.~\ref{fig:gnn_results_val} was on the order of $30,000$ CPU hours on Intel Xeon Gold 6230 CPUs. We ran into RAM limitations for $\beta=3$ experiments after $N=10^7$, and could not increase $N$ further due to this.

\textbf{Turning the knob and trivial syndromes} During training, trivial syndromes (corresponding to no error) were discarded and counted towards model accuracy; for $\beta \in (1, 2, 3)$ the fraction of trivial examples was approximately $(75\%, 57\%, 43\%)$. This suggests a simpler mechanism to explain the effectiveness of turning the knob: large $\beta$ generates more nontrivial examples for training. To control for this, Fig.~\ref{fig:gnn_nontrivial} rescales Fig.~\ref{fig:gnn_results_val} from the main text to report only the number nontrivial data. This result shows that models trained with larger $\beta$ continue to outperform models trained using smaller $\beta$, even when controlled for the same number of nontrivial examples seen by each model. This implies that the performance improvement from turning the knob is partially due to nontrivial examples in the training dataset$D_N[\beta p]$ being better for training than $D_N[p]$, and the theory of sample importance provides a better explanation for this compared to simply counting the number of nontrivial syndromes.

\end{document}